\documentclass[12pt, onecolumn]{IEEEtran}
\usepackage{amsmath,amssymb,eucal,graphicx}
\usepackage{pst-plot}
\usepackage{epsfig}
\usepackage{amsthm}
\usepackage{amsfonts}
\usepackage{epsfig}
\usepackage{float}
\usepackage{pstricks}
\usepackage{color}
\usepackage{mathcomSTEP,stmaryrd}
\usepackage{amsmath,subfigure}
\usepackage{setspace}
\usepackage{subfigure}

\newtheorem{definition}{Definition}

\theoremstyle{remark}

\begin{document}

\title{
Energy Efficient Cooperative Strategies for Relay-Assisted Downlink Cellular Systems\\
Part II: Practical Design
%
}

\author{%
\authorblockN{%
Stefano~Rini\authorrefmark{1}\authorrefmark{2},
Ernest~Kurniawan\authorrefmark{2},
Levan~Ghaghanidze\authorrefmark{1},
and
Andrea Goldsmith\authorrefmark{2}\\
}
\authorblockA{%
\authorrefmark{1}
Technische Universit\"{a}t M\"{u}nchen, Munich, Germany\\
E-mail: \{stefano.rini,levan.ghaghanidze\}@tum.de} \\
\authorblockA{%
\authorrefmark{2}
Stanford University, Stanford, CA, USA \\
E-mail: ernestkr@stanford.edu, andrea@wsl.stanford.edu }
}

\maketitle

\begin{abstract}

In a companion paper \cite{RiniEnergyPartI13}, we present a general approach to evaluate the impact of cognition in a downlink cellular system in which multiple
relays assist the transmission of the base station.
This approach is based on a novel theoretical tool which produces transmission schemes involving rate-splitting, superposition coding and interference decoding
for a network with any number of relays and receivers.
%
%
This second part focuses on a practical design example for a network in which a base station transmits to three receivers with the aid of two relay nodes.
For this simple network, we explicitly evaluate the impact of relay cognition and precisely characterize the trade offs between the total energy consumption
and the rate improvements provided by relay cooperation.
These closed-form expressions provide important insights on the role of cognition in larger networks and highlights interesting interference management strategies.
%
%
%

We also present a numerical simulation setup in which we fully automate the derivation of achievable rate region for a general relay-assisted downlink
cellular network.
Our simulations clearly show the great advantages provided by cooperative strategies at the relays as compared to the uncoordinated scenario under varying
channel conditions and target rates.
These results are obtained by considering a large number of transmission strategies for different levels of relay cognition and numerically determining
one that is the most energy efficient.
The limited computational complexity of the numerical evaluations makes this approach suitable for the optimization of transmission strategies for larger networks.

\end{abstract}

\section{Introduction}\label{chap:introduction}

Wireless relay nodes hold the promise of drastically increasing both the energy and spectral efficiency of future cellular systems \cite{pabst2004relay}.
The demand of transmission rates has increased tremendously between the third and fourth generations
of mobile networks due to new functionalities and services provided by handset devices.
Rate gains of this order cannot be achieved without increasing bandwidth or the density of the network and relay nodes represent a
low-cost way of achieving such increase.
Despite the relevance of relays in practical systems, it is not yet clear what system architecture is the most energy efficient and what advantages are provided
by cognition in this scenario.
%
%
%
%
%
%


The architecture of relay-assisted downlink cellular system in LTE-A in presented in Fig.  \ref{fig:caseexample}(a)
: the system is comprised of a base station
which is interested in communicating to multiple receivers with the aid of the relay nodes.
The set of communications between the base station and the relay nodes is termed a \emph{relay link} while the set between relay nodes and receives is termed an \emph{access link}.
We consider the case where no direct link between the base station and receivers exists: this case can be easily obtained by considering an additional relay which is connected to the base station with an infinity capacity channel.
In the relay link, transmissions take place over frequency separated channels and are thus non interfering.
In the access link, instead, transmissions take place over the same frequency band and therefore are self interfering.
When relays cooperate, the access link is analogous to a multi-terminal cognitive channel in which transmitting node are able to partially coordinate their transmissions.

The difficulty in designing efficient communication strategies for this system lies in the intrinsic complexity of the communication strategies that can be implemented with even a moderately small number of relays and receivers.
Possible ways to circumvent this obstacle is to either consider simple communication protocols for each relay or to assume that relays operate in a uniform
manner which does not depend on the network conditions.
Two such examples are the so called \emph{relay selection} strategy \cite{LueLinkFailure05,LinCooperativeRegions04,LiRelaySearch05} and a generalization of the
amplify-and-forward or decode-and-forward strategies for the relay channel \cite{zhao2006improving}.

In \cite{RiniEnergyPartI13} we propose a different approach: we generalize the information theoretical derivation of achievable schemes to a network with any number of relays and receivers and obtain the achievable rate region from the analysis of the error probability for this general scenario.
This approach is fundamentally different from the previous approaches as it considers all possible ways in which the base station can distribute messages to
the relay nodes through cognition and as well as all possible cooperation strategies that can be implemented at the relay.
In particular we focus on the role  of partial, or unidirectional, cooperation at the relay which allows for coordinated transmission strategies which have been studied in the context of cognitive channels such as the Cognitive InterFerence Channel (CIFC) \cite{devroye2005cognitive} and the Interference Channel with a Cognitive Relay (IFC-CR) \cite{Sahin_2007_1}.

As a result, the number and the complexity of the scheme considered by this approach has no comparison with other approaches.
Moreover, the generation and the evaluation of the schemes has a low computational complexity  and it is thus possible to analyze large networks and determine the
optimal level of relay cooperation and the most energy efficient transmission strategy.

\subsection{Literature Overview}\label{sec:motivation}

In the literature, different relay-assisted downlink  systems have been proposed and studied.
A conceptually simple and yet powerful approach is \emph{relay selection} \cite{LueLinkFailure05,LinCooperativeRegions04,LiRelaySearch05},
i.e. selecting the relay which  allows for the fastest data transmission between the base station and receivers.
Perhaps surprisingly,  this scheme provides the same cooperative diversity as more complicated strategies \cite{BletsasSimpleCooperativeDiversityMethod06}.
Space-time coding for relay-assisted transmissions has been considered in \cite{stefanov2005cooperative} and it has been shown to achieve full cooperation diversity at the cost of an increased complexity.
The sum rate optimization for amplify-and-forward multi-antenna relay networks is considered in \cite{lee2008achievable}, where the ``power iterative algorithm'' for achieving optimal beamforming is proposed. Achievable rate regions in two-way relay communications is discussed in \cite{rankov2006achievable}. Combined strategies using amplify-and-forward, decode-and-forward, compress-and-forward with superposition coding and other encoding and decoding techniques are compared for different scenarios.

In the context of energy efficiency, virtual MIMO systems have been considered to exploit the broadcast nature of the medium and distribute messages from the base station to the relay nodes.
In this case it is possible to analyze the performance of a system where relay nodes opportunistically decode the transmissions that can be overheard over
the network.
When multiple relays are able to decode multiple transmissions, they are then able to act as a virtual MIMO system in the next hop.
This can be incorporated into multi-hop transmission schemes to study the energy efficiency as in \cite{WenqingVirtualMIMOprotocol05} and \cite{CosoVirtualMIMO06}.
Energy efficiency for relay-assisted LTE-A networks is studied in \cite{FantiniE3F11,FantiniEnergyEfficiency11} where it is shown that a multi-cast cooperative scheme in relay networks results in more efficient energy consumption compared to a
conventional two-hop approach.
This approach does not consider the cooperation among multiple relays and considers only limited transmission strategies, but it is a first approach to this problem.
Another approach to energy efficiency analysis of cooperative relaying is considered in \cite{Energy:Xiao}-\cite{Energy:Madan}: here an opportunistic scheme is proposed, which  relies on the channel state information (CSI) knowledge to decide which subset of the relay nodes cooperate.
%
%

\subsection{Contributions}

Previous literature focuses on the case of either no relay cooperation or full relay cooperation. We  focus here the intermediate case of partial and unidirectional transmitter cooperation.
In information theory, these forms of cooperation are referred to as cognition as in the CIFC \cite{devroye2005cognitive} and the IFC-CR \cite{Sahin_2007_2}.
The two channels idealize  the ability of some transmitter to learn the message of the other user thanks to the broadcast nature of the wireless medium.
%

In a companion paper \cite{RiniEnergyPartI13}, we propose a new theoretical framework to design optimal transmission strategies for the case where each relay has knowledge of multiple messages.
We are able to derive the achievable rate region of any scheme transmission scheme at the relay involving rate-splitting, superposition coding and interference decoding.
The schemes to be considered on the access link depend on the message decoded at the relay nodes, which we refer to as \emph{level of cognition}.

Although very powerful, this approach is too general  to provide  much insight on the important features of energy efficient transmission schemes.
In the following we specialize the approach of \cite{RiniEnergyPartI13} to the network where a base station communicates to three receivers with the aid of two
relay nodes.
For this specific network we provide closed-form expression for the achievable rate regions which provide significant insights on the tarde off between cooperation
and interference management at the relays.
We also present a simulation setup which shows how the theoretical results in \cite{RiniEnergyPartI13} can be efficiently implemented.
This shows that the proposed approach can indeed be used to analyze larger networks.

The contributions in the paper can be summarized as follow:
%
%
%
%
%

\noindent
\textbf{Explicit Characterization for Power Consumptions: }
Using the results in \cite{RiniEnergyPartI13}, we derive explicit characterizations for the power consumption of several transmission strategies
and provide insights on the tradeoff between power consumption at the base station and cooperation at the relays.
By comparing the performance of different schemes, we show that cooperation is not only useful in increasing the
power efficiency relative to the uncoordinated case but indeed is necessary to attain high data rates.

\bigskip

\noindent
\textbf{Numerically Determine Optimal Strategies}
We automate the derivation of achievable rate regions in \cite{RiniEnergyPartI13} and through computer simulation we determine the most energy efficient
transmission strategies for a number of channel realizations.
In each case, we consider a number of possible achievable strategies in the order of the thousands for the channel model under consideration.
The minimization of the power consumption necessary to achieve a target rate for each scheme can be performed with almost linear complexity.
Indeed the fact that our approach is able to consider a large number of transmission schemes and optimize each scheme with reasonable complexity makes
it very suitable for larger networks.


\bigskip

\subsection{Paper Organization}

Section \ref{sec:model} introduces the network model under consideration: the relay-assisted downlink cellular system in with two relays and three receivers.
The problem of energy minimization this network is introduced in Sec. \ref{sec:Problem Description and an Illustrative Example}.
Lower bounds to the energy efficiency are provided in Sec. \ref{sec:Lower Bounds to the Energy Efficiency}.
Explicit characterizations for the network topology under consideration are provided in Section \ref{Sec:Considerations on a Simple Example}.
The results of the numerical optimization are provided in Section \ref{sec:Numerical Evaluations}.
Finally, Section \ref{sec:Conclusion} concludes the paper.

\subsection{Notation}

In the remainder of the paper we adopt the following notation:

\begin{itemize}
  \item variables related to the Base Station (BS) are indicated with the superscript $\rm BS$, moreover $i$ is the index related to BS,

  \item variables related to the Relay Nodes (RN) are indicated with the superscript $\rm RN$, moreover $j$ is the index related to RNs,

  \item variables related to the Receivers (RX) are indicated with the superscript $\rm RX$, moreover $z$ is the index related to RXs,

  \item $\Ccal(\Sigma)=1 / 2 \log \lb | \Sigma \Sigma^H +\Iv|\rb$ where $X$ is a vector of length $k$ of jointly Gaussian random variables and $|A|$ indicates the determinant of $A$,

  \item $A_{ij}$ element of the matrix $A$ in row $i$ and column $j$,
\end{itemize}

\section{Channel Model}
\label{sec:model}

\begin{figure}[!htb]
    \centering
    \includegraphics[width=0.75\textwidth]{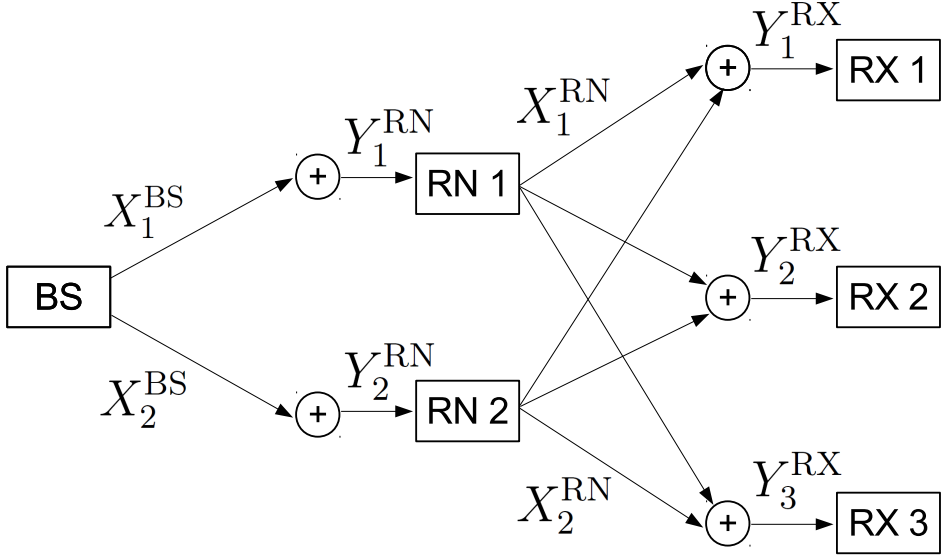}
    \vspace{-3.3 cm}
    \caption{ The relay-assisted downlink cellular system in with two relays and three receivers.  }
    \label{fig:channel_model_2X3}
\end{figure}

In the following we consider a relay-assisted downlink cellular system with two relays and three receivers.
In this model, also depicted in Fig. \ref{fig:channel_model_2X3}, the  Base Station (BS) is interested in communicating to three receivers (RXs)
with the aid of two Relay Nodes (RNs).
This model is a special case of the general model in \cite{RiniEnergyPartI13} in which $N_{\rm RN}=2$ and $N_{\rm RX}=3$, in particular
each RX $z\in \{1,2,3\}$ is interested in the message $W_z$ at rate $R_z$ which is known at the BS and is to be transmitted reliably
and efficiently to RXs through the RNs~1 and~2.
%
The channel inputs at the Base Station (BS) are $X^{\rm BS}_1$ and $X^{\rm BS}_2$ each subject to the power constraint
is $E[|X^{\rm BS}_1 |]+E[|X^{\rm BS}_2 |] \leq P^{\rm BS}$.
The channel outputs at Relay Node(RN) one and two
\eas{
Y_1^{\rm RN} & = d_{11} X^{\rm BS}_1  + Z^{\rm RN}_1 \\
Y_2^{\rm RN} & = d_{22} X^{\rm BS}_2  + Z^{\rm RN}_2.
}
In a similar fashion, the channel input at the RNs are $X^{\rm RN}_1$ and $X^{\rm RN}_2$ subject to the
power constraints $E[|X^{\rm RN}_1 |]\leq P^{\rm RN}_1$ and $E[|X^{\rm RN}_2 |]\leq P^{\rm RN}_2$.
The channel outputs at the RXs are
\eas{
Y_1^{\rm RX} & = h_{11} X^{\rm RN}_1  + h_{12} X^{\rm RN}_2  + Z^{\rm RX}_1 \\
Y_2^{\rm RX} & = h_{21} X^{\rm RN}_1  + h_{22} X^{\rm RN}_2  + Z^{\rm RX}_2 \\
Y_3^{\rm RX} & = h_{31} X^{\rm RN}_1  + h_{32} X^{\rm RN}_2  + Z^{\rm RX}_3.
}
Each noise term $Z$ has variance one and the channel coefficient can take any complex value.
A graphical representation of the channel is provided in Fig. \ref{fig:channel_model_2X3}.

The transmission between the BS and the RNs as well as the transmission between the RNs and the RXs takes place over $N$ channel transmissions.
Each message $W_z$ is uniformly distributed in the interval $[1 \ldots 2^{N R_z}]$.
Let $W$ indicate the vector containing all the messages to be transmitted, i.e. $W=[W_1 \ W_2 \ W_3]$ and $R$
 the vector containing the rate of each message, i.e. $R=[R_1 \ R_2 \ R_3]$.
Additionally let $W_j^{\rm RN}$  be the set of messages decoded at relay node $j \in \{1,2\}$ and define $W^{\rm RN}=[W_1^{\rm RN} \ W_2^{\rm RN}]$.
A transmission on the relay link is successful if there exists an encoding function at the BS and a decoding function at each RN such that each relay can successfully decode the message in $W_j^{\rm RN}$ with high probability.
Similarly, a transmission on the access link is successful if there exists an encoding function at each RN and a decoding function at each RX such that
each receiver $z \in \{1,2,3\}$ can decode the message $W_z$ reliably.
More formally, let $\Wh_{z}^{RN_j} $ be the estimate of $W_z$ at relay $j$ and $\Wh_{z}$ the estimate of $W_z$ at receiver $z$ over $N$ channel transmissions, then a communication error occurs when there exist
$\Wh_{z}^{RN_j} \neq W_z$  or $\Wh_{z} \neq W_z$
for some noise realization over the relay link or the access link.

A rate vector
$R$
is said to be achievable if,
for any $\epsilon > 0$, there is an $N$ such that
\begin{align*}
\max_z \max_{W^{\rm RN}_j} \ \Pr \lsb  \Wh_{z}^{RN_j} \neq \Wh_{z} \neq W_{z} , \rsb \leq \epsilon.
\end{align*}
Capacity is the closure of the union of the sets of achievable rates.

In the following we consider the problem of minimizing $E_{\rm TOT}$, the total energy required to achieve a rate vector $R=[R_1 \ R_2 \ R_3]$ defined
as
\eas{
E_{\rm TOT} & = \f {P_{\rm TOT}} { R_1 + R_2 + R_3},  \quad \quad P_{\rm TOT} & = P^{BS} + \sum_{j=1}^{3} \mu_j \f{ P^{\rm RN}_{j} } {R_j}.
%
}

\section{Problem Description}
\label{sec:Problem Description and an Illustrative Example}

In this section we provide some initial insight on the role of cooperation for the network model in Sec. \ref{sec:model}.
We are interested in characterizing  the energy advantages provided by cooperation and cognition when employing superposition coding, interference decoding and rate-splitting.
When the base station distributes the same message to multiple RNs, the access link becomes cognitive channel in the sense of the
Cognitive InterFerence Channel (CIFC) and the InterFerence Channel  with a Cognitive Relay (CIFC-CR), that is, transmitting nodes have partial or
unidirectional knowledged of each others' messages and can thus perform partially coordinated transmissions.
Transmitter cooperation has been shown to provide substantial improvement on the rate performance \cite{rini2012inner} as well as suggesting new and
surprising cooperative transmission schemes \cite{rini2010new}.
In order to obtain the performances of cognitive channels though, it is necessary to distribute the same message to multiple RNs
which consumes more power in the relay link.
Indeed we are interested in characterizing the tension between increase of energy in the relay link and the benefits provided by cognition in the access link .
In general, there exists an optimal level of cognition at the relay nodes for which the additional energy consumption at the relay link allows for the
most beneficial performance improvements in the access link.
Our aim is to quantify the amount of RN cooperation that is needed to attain the largest energy efficiency and determine what coding choices attain it.
%
%

\begin{figure}[!htb]
    \centering
    \includegraphics[width=0.75 \textwidth]{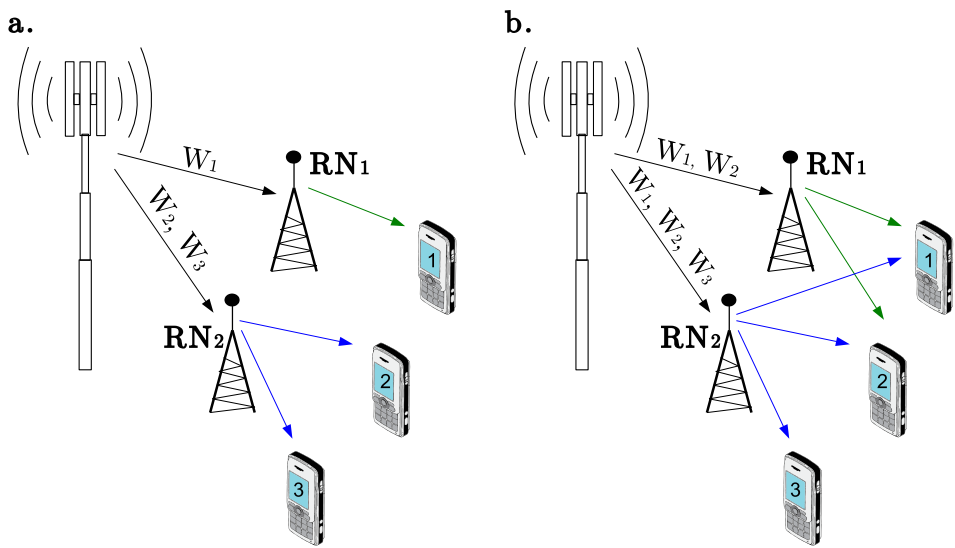}
    \vspace{-1.2 cm}
    \caption{An example comparison of using cooperative transmission strategy and transmitting without cooperation.
    %
    }
    \label{fig:caseexample}
\end{figure}

Let us consider the scenario in which the BS distributes each message to only one RN, as illustrated in Fig. \ref{fig:caseexample}.A.
In this case, the power utilization at the BS is minimal but no cooperation between the relays is possible as no message is known at both.
The lack of cooperation among RNs corresponds, in general, to a higher power consumption on the access link as there is no diversity gain at the receivers: when multiple RNs send the same signal, we obtain a combining gain at the receiver.
Cooperation provides some additional advantages beyond combining: by knowing the same message, RNs can each superimpose the codewords of other users over
the codeword of the common message.
This, in general, achieves larger energy savings than sending each codeword separately.
Consider now the message allocation at the RNs in Fig. \ref{fig:caseexample}.B: since message  $W_1$ and $W_2$ are distributed to both RNs, it is possible for the RNs to cooperate in transmitting these messages and the codeword to transmit $W_3$ can be superimposed over the common codewords.
The transmission of a message to multiple RNs, however, utilizes more power on the relay link and the advantages attained on the access link must be weighted against
this additional power consumption.

The aim of the paper is to investigate the tradeoff between cooperation and power consumption when considering classical information theoretical achievable strategies. We attempt to identify the message allocation at the relay nodes and the relay cooperation strategy which allows for the lowest overall power consumption.
%

In the network of Sec.  \ref{sec:model}, the possible levels of relay cognition are:

\smallskip
\noindent
{\bf $\bullet$ No cooperation: each message is known only at one RN.}
The RNs code independently and the RN that knows both message can apply superposition coding among these codewords.

This scenario minimizes the power consumption at the relay link but no gain from cooperation is possible in the access link.

\smallskip
\noindent
{\bf $\bullet$ Partial Cooperation I: a message is known at both RN. }

The codeword of the common message acts as a base codewords on top of which the other two codewords can be superimposed.
When superposition coding is applied, the RXs decoding the top codewords must also decode the bottom one:
 this imposes additional rate constraint on the bottom codeword but
improves the energy efficiency in sending the top codewords.

\noindent
{\bf $\bullet$ Partial Cooperation II: two messages are known at both RN.}

The codeword of the two common messages can be superimposed one on top of the other and act as base codewords for the message known only at one RN.

This improves on the energy gains  offered by superposition coding with respect to the previous scenario but consumes additional power on the relay link.

\noindent
{\bf $\bullet$  Full Cooperation: all messages are known at all RNs.}
In this case the power consumption at the base station is maximal but the RNs can cooperate on the transmission of every codeword.

\begin{figure}[ht]
\centering
\subfigure[Different superposition coding schemes for different levels of cooperation.]{
\includegraphics[width=.47 \textwidth]{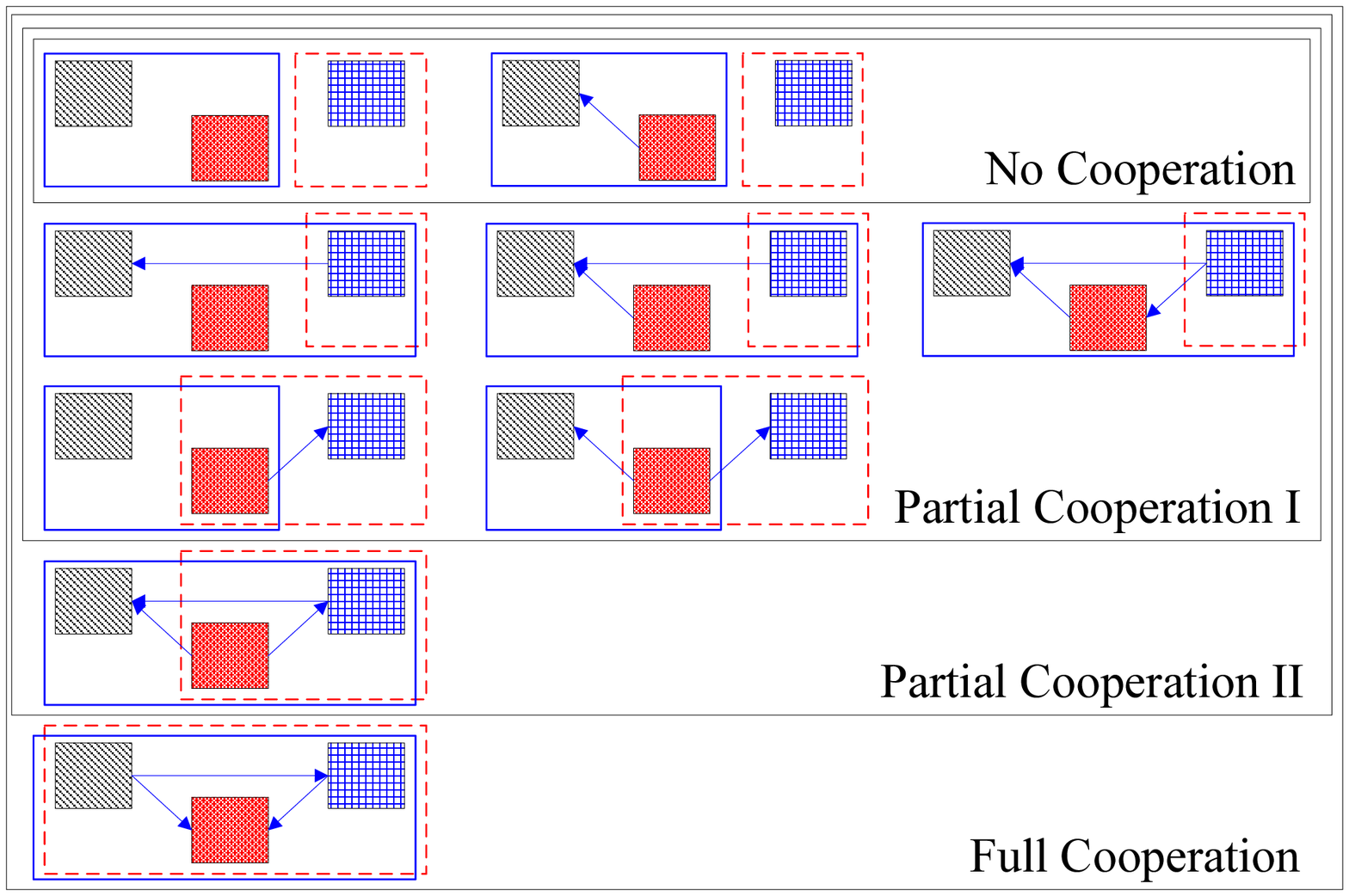}
    \label{fig:schemesdesc}
}
\subfigure[Different superposition coding schemes with no, one, two or three superposition coding steps.]{
\includegraphics[width=.47 \textwidth]{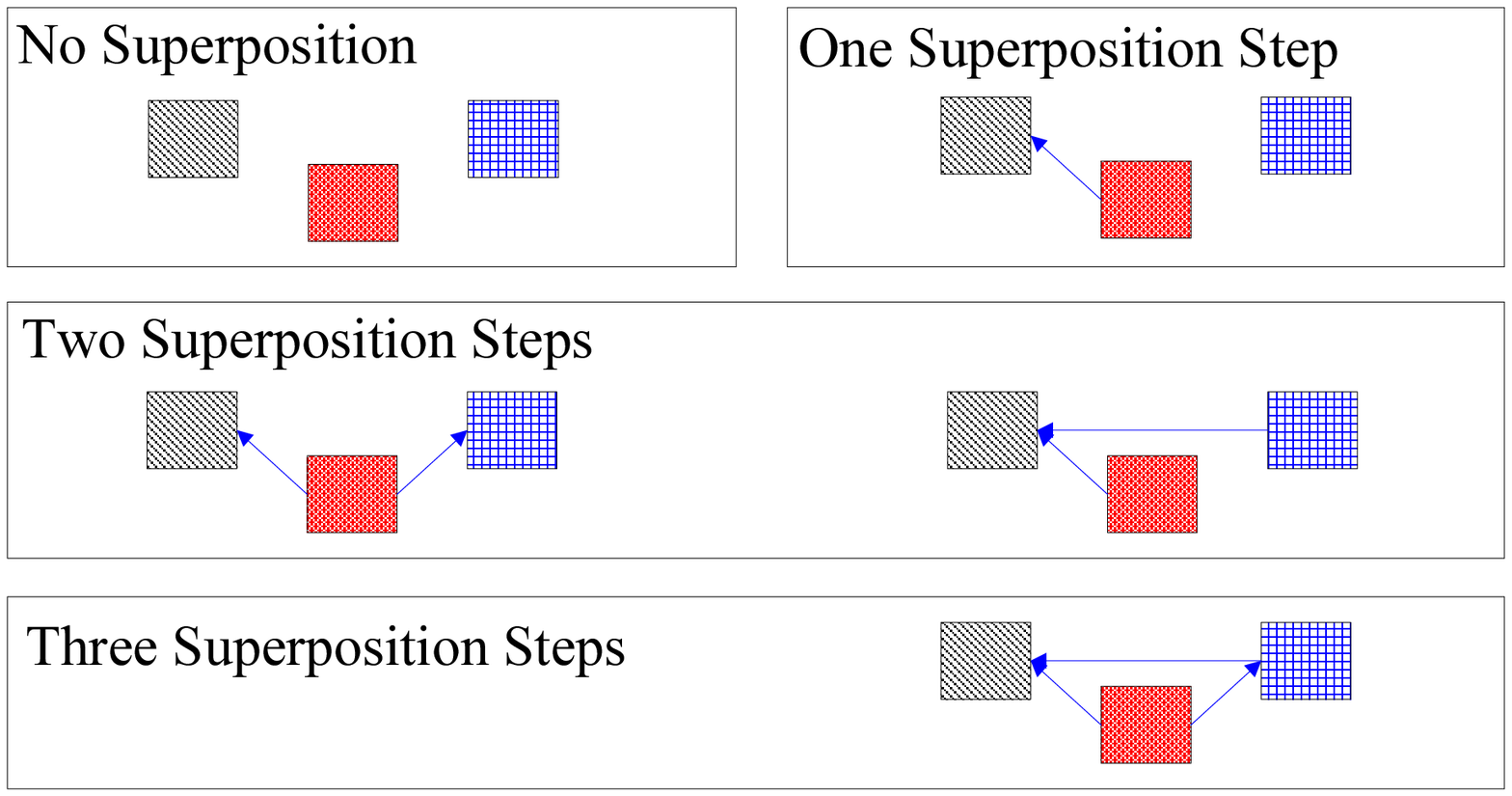}
    \label{fig:superdesc}
}
\caption{Cognition and cooperation for the relay-assisted downlink cellular system in with two relays and three receivers.}
\end{figure}

\noindent
\bigskip
The different transmission strategies corresponding to the different level of cooperation are depicted in Fig. \ref{fig:schemesdesc}.
Each hatched box represents the codeword of a message: codewords are not labeled to account for the symmetry of the possible $W_j^{\rm RN}$.
The boxes encircled by the solid blue line are known at one RN while the boxes encircled by a dotted red line by the other RN
(no labeling is used to account for symmetry).

Since the different cases are listed with increasing number of shared messages, at each step the transmission strategies of the previous cases are also viable.

Arrows connecting boxes represent the superposition coding: the orientation is from the bottom codewords toward the top one.

The decoding of the top codeword requires also the decoding of the bottom codewords. For this reason, superposition coding restricts the rate of the base codeword to the highest rate that can be reliably decoded by all the receivers which must decode this codeword.
Consequently applying superposition coding does not always result in a rate advantage: this is the case only when the decoding of the base codeword at a non-intended receiver does not limit the achievable rates at the intended one.

\smallskip
To distinguish among possible application of superposition coding, we can classify the possible encoding strategies depending on the number of superposition coding steps they involve: one, two or three.
The possible encoding strategies in each case are depicted on Fig. \ref{fig:superdesc}.

%
%
%
%
Different schemes require at least a certain level of cooperation among relays and the same scheme can be implemented for different levels of cooperation.
This classification uniquely identifies the number of interfering codeword that a receiver is required to decode.

\section{Lower Bounds to the Energy Efficiency}
\label{sec:Lower Bounds to the Energy Efficiency}


In this section we specialize the results in \cite[Th. V.1.]{RiniEnergyPartI13}, \cite[Th. V. 2.]{RiniEnergyPartI13} and \cite[Lem. V. 3.]{RiniEnergyPartI13} to the channel model in Sec. \ref{sec:model}.
Although simple, this example provides the reader with the basic intuition on the structure and role of these bounds.

\begin{lem}{\bf Relay Link Capacity for the Two Relays and Three Receivers Case}
\label{lem:Relay Link Capacity for the Two Relays and Three Receivers Case}
For a given message allocation $W^{\rm RN}$ the capacity of the relay link  for the channel model in Sec. \ref{sec:model} is
\eas{
\sum_{ z \in W_1^{\rm RN} } R_z & \leq \Ccal( |d_{11}|^2 P_1^{\rm BS})\\
\sum_{ z \in W_2^{\rm RN} } R_z & \leq \Ccal( |d_{22}|^2 P_2^{\rm BS}),
}{\label{eq:Relay Link Capacity for the Two Relays and Three Receivers Case}}
for any $P_1^{\rm BS}$ and $P_2^{\rm BS}$ such that $P_1^{\rm BS}+P_2^{\rm BS}=P^{\rm BS}$.
\end{lem}

Now for the access link outer bounds we have

\begin{lem}{\bf Access Link Outer Bound for the Two Relays and Three Receivers Case}
\label{lem:Access Link Outer Bound for the Two Relays and Three Receivers Case}

Given a message allocation   $W^{\rm RN}$, an outer bound to the achievable rates on the access link for the channel in Sec. \ref{sec:model} is
\eas{
R_1 & \leq  \Ccal \lb H_{11} A_{1 1}+H_{12} A_{2 1}\rb \\
R_2 & \leq  \Ccal \lb H_{21} A_{1 2}+H_{22} A_{2 2}\rb \\
R_3 & \leq  \Ccal \lb H_{31} A_{1 3}+H_{32} A_{2 3}\rb \\
R_1 + R_2 & \leq  \Ccal \lb  \lsb \p {H_{11} & H_{12}  \\ H_{21} & H_{22}} \rsb  \lsb \p {A_{11} & A_{12}  \\ A_{21} & A_{22} }\rsb\rb \\
R_1 + R_3 & \leq  \Ccal \lb  \lsb \p {H_{11} & H_{12}  \\ H_{31} & H_{32}} \rsb  \lsb \p {A_{11} & A_{13}  \\ A_{21} & A_{23} }\rsb\rb \\
R_2 + R_3 & \leq  \Ccal  \lb  \lsb \p {H_{21} & H_{22}  \\ H_{31} & H_{32}} \rsb  \lsb \p {A_{12} & A_{13}  \\ A_{22} & A_{23} }\rsb\rb \\
R_1 + R_2 + R_3 & \leq  \Ccal  (H A),
}{\label{eq:Access Link Outer Bound for the Two Relays and Three Receivers Case}}
%
%
union over all the possible matrices $A$ as defined in \cite[eq. (27)]{RiniEnergyPartI13}.
\end{lem}
%
%

From Lem. \ref{lem:Relay Link Capacity for the Two Relays and Three Receivers Case} and Lem.  \ref{lem:Access Link Outer Bound for the Two Relays and Three Receivers Case} we obtain a lower bound on the energy consumption by using the fact that determining capacity is the dual problem of determining power efficiency.

\begin{lem}{\bf Energy Consumption Lower Bound}
\label{lem:Energy Efficiency Lower Bound for the Two Relays and Three Receivers Case}

A lower bound to energy required for achieving the rate vector $R$ in the channel model in Sec. \ref{sec:model} is obtained as the minimum over  $W^{\rm RN}$ defined as
\eas{
W_1^{\rm RN} , W_2^{\rm RN} & \subseteq \{1,2,3\} \\
W^{\rm RN} & = \lcb W_1^{\rm RN} , W_2^{\rm RN} \rcb,
}
of $P^{\rm BS}$ and $[P_1^{\rm RN} \ldots P_{N_{\rm RN}}^{\rm RN}]$ such that $R$ is on the boundary of
\eqref{eq:Relay Link Capacity for the Two Relays and Three Receivers Case} and \eqref{eq:Access Link Outer Bound for the Two Relays and Three Receivers Case}
respectively.

\end{lem}
%
%

\section{Achievable Schemes}
\label{Sec:Considerations on a Simple Example}

In this section we provide some closed-form expressions of achievable rate region which provide concrete examples of the tradeoffs between cooperation and power consumption that emerge in the set up we consider.
This section also sets the stage for the numerical simulations is Sec.  \ref{sec:Numerical Evaluations} where some of the schemes presented here are simulated numerically.

In particular we consider four schemes
\begin{itemize}
  \item \textbf{Non-cooperative scheme with two active relays:} this scheme minimizes the power consumption on the relay link by distributing the one message to one relay and two messages to the other.
  \item \textbf{Non-cooperative scheme with two active relays and interference decoding:} even when non cooperative schemes are implemented, interference decoding can be used to boost the energy efficiency
  \item \textbf{Partially cooperative scheme:} in which the relays cooperate in transmitting at least one message
  \item \textbf{Partially cooperative scheme with interference decoding:} which combines both cooperative gains and interference decoding gains.
  %
%
\end{itemize}
The schemes above are taken without rate-splitting, so that $\Gamma$ is a binary matrix.
In the following,  to avoid having to specify the rate-splitting matrix in each different scenario, we adopt the notation $U_{\iv \sgoes \jv}( W_z)$ to indicate that the message $W_z$
is embedded is the codeword $U_{\iv \sgoes \jv}^N$.

We again focus on the simple channel model in Sec. \ref{sec:model} and additionally focus on the symmetric case in which
\ea{
H=
\begin{bmatrix} 1&b \\ a&a \\ b&1 \end{bmatrix},
\label{eq:symmetry simple case}
}
for $a,b \in \Rbb^+$ and
\ea{
d_{11}=d_{22}=1,
\label{eq:symmetry simple case relay link}
}
%
%
%
With this specific choice of channel parameters, we  have that
\begin{itemize}
  \item by fixing $a=0$, the access channel reduces to an interference channel,
  \item by fixing $b=0$, no communication is possible between RN~1 and RX~3 and between RN~2 and RX~1,
  \item when one of the relay has knowledge of all messages and the other none, the channel reduces to a degraded Gaussian Broadcast Channel (BC),
  \item when $a=0$ and RNs know independent messages, the access channel reduces to an interference channel,
  \item when $a=0$ and a relay node knows the message of the other, the access channel reduces to a cognitive interference channel,
\end{itemize}

\begin{figure}[h]
\centering
\includegraphics[width=0.8\textwidth]{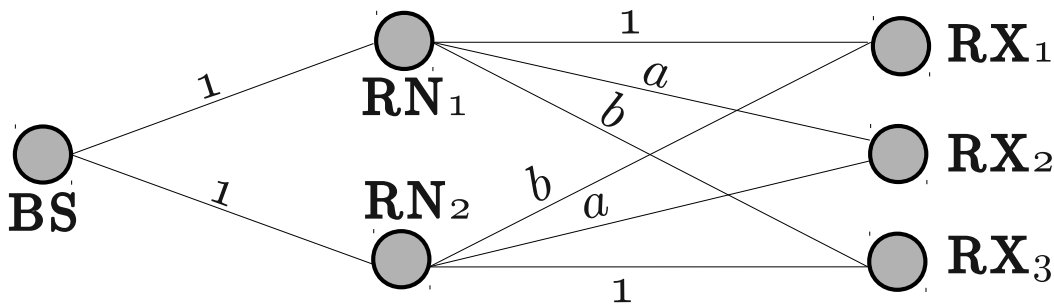}
\vspace{ -0.5 cm}
\caption{
The symmetric channel model considered in Sec. \ref{Sec:Considerations on a Simple Example}. }
\label{fig:simsetup_a_b}
\end{figure}

A graphical representation of the channel under consideration is provided in Fig. \ref{fig:simsetup_a_b}.

%
%
We also focus on the symmetric rate problem, that is we fix $R_{\rm sym}=R_1=R_2=R_3$.
%
By choosing a highly symmetric model and symmetric rates, we make it possible to derive simple explicit characterizations which lend themselves to intuitive interpretations
and a very simple graphical representation.

\begin{figure}[h]
\centering
\includegraphics[width=\textwidth]{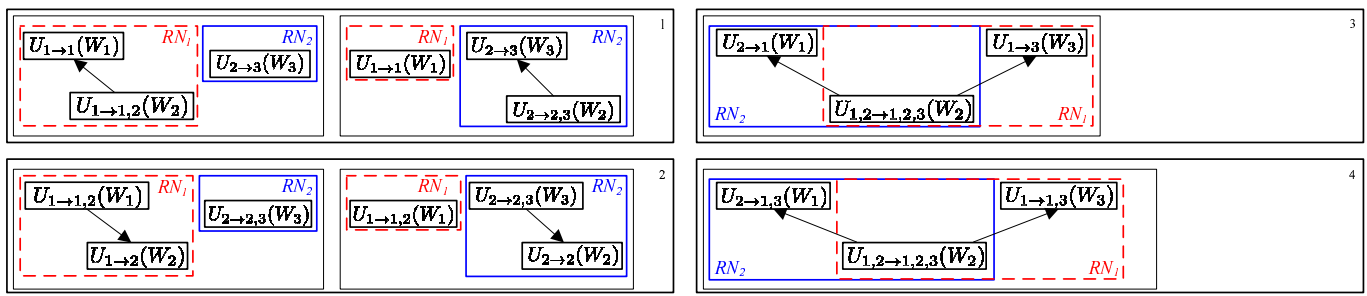}
\vspace{ -0.5 cm}
\caption{
The four CGRAS presented in Sec. \ref{Sec:Considerations on a Simple Example}.
}
\label{fig:Schemes}
\vspace{- 0.5 cm}
\end{figure}

\subsection{Non-cooperative scheme with two active relays}
\label{sec:Non-cooperative scheme with two active relays}

The power consumption at the BS is minimized when transmitting one message to a RN and two messages to the other RN, in  which case $P^{\rm BS}= 16^{R_{\rm sym}}+4^{R_{\rm sym}}-2$.
In this case, superposition coding can be applied at the relay node that knows two messages: the order of superposition of the codeword is chosen according to the SINR toward the two RXs for which the messages are intended.
%
%
This scheme is in general efficient also on the access link but it is interference limited at the decoders which treat the interference as noise.
For this reason, it may not be possible to find a power allocation which achieves a given target rate.
%

Let's consider the case in which RN~1 has knowledge of $W_1$ and RN~2 has knowledge of $W_2$ and $W_3$ (other  message combinations are easily obtained in a similar manner and also considering the symmetry of the channel).
This assignment is advantageous in the case $a < 1$ and $b <1$  in which case it is more convenient to send $W_1$ through RN~1 instead of RN~2 and $W_3$ through RN~2.
Also, since $a < 1$, RN~2 it is advantageous to superimpose the codeword for $W_3$ over the one for $W_3$.
%

This corresponds to the CGRAS number~1 in Fig. \ref{fig:Schemes} (same as CGRAS number~3 in Fig. \ref{fig:rate01}).
With  this coding choice the following rate region is achievable
\eas{
R_1 & \leq \Ccal \lb \frac{P_{11}}{1 + b^2 (P_{22} + P_{23}) } \rb
\label{eq:schemde power efficiency relay link R1} \\
R_2 & \leq \Ccal \lb \frac{a^2 P_{22}}{1 + a^2 P_{11} + a^2 P_{23}} \rb
\label{eq:schemde power efficiency relay link R2} \\
R_2 + R_3 & \leq \Ccal \lb \f{ P_{22} + P_{23}} {1+ b^2 P_{11}}\rb
\label{eq:schemde power efficiency relay link sum rate} \\
R_3 & \leq \Ccal \lb \frac{P_{23}}{1 + b^2 P_{11}} \rb,
}{\label{eq:schemde power efficiency relay link}}
with the assignment
\eas{
X_1 & =  \sqrt{P_{11}} U_{1 \sgoes 1}(W_1) \\
X_2 & =  \sqrt{P_{22}} U_{2 \sgoes 23 }(W_2) + \sqrt{P_{23}} U_{2 \sgoes 3}(W_3),
}
which has a total power consumption at the access link of
\ea{
P^{\rm RN}_1+P^{\rm RN}_2= P_{11} + P_{22} + P_{23}.
}

When $a\geq 1 \geq  b$, the sum rate bound \eqref{eq:schemde power efficiency relay link sum rate} is never redundant which means that the  region in \eqref{eq:schemde power efficiency relay link} has two Pareto optimal corner points.
With the assumption, $R_{\rm sym}=R_1=R_2=R_3$, the required transmit power can then be calculated as:
\eas{
P_{11} &\geq \Ccal^{-1}(R_{\rm sym}) \lb 1 + b^2 (P_{22} + P_{23}) \rb
\label{eq:schemde power efficiency relay link min powers P11}\\
P_{22} &\geq \Ccal^{-1}(R_{\rm sym}) \lb a^{-2} + P_{11}+P_{23}\rb
\label{eq:schemde power efficiency relay link min powers P22} \\
P_{22} + P_{23} &\geq \Ccal^{-1}(2R_{\rm sym}) \lb 1 + b^2 P_{11}\rb
\label{eq:schemde power efficiency relay link min powers P22 P23}\\
P_{23} &\geq \Ccal^{-1}(R_{\rm sym})\lb 1 + b^2 P_{11}\rb,
\label{eq:schemde power efficiency relay link min powers P23}
}{\label{eq:schemde power efficiency relay link min powers}}
for $\Ccal^{-1}(x)=4^x-1$.

When combining equations \eqref{eq:schemde power efficiency relay link min powers P11} and \eqref{eq:schemde power efficiency relay link min powers P22 P23} we obtain
\ea{
P_{11} \lb 1 - b^4 \Ccal^{-1}(2 R_{\rm sym})\Ccal^{-1}(R_{\rm sym}) \rb & \geq \Ccal^{-1}(R_{\rm sym}) \lb 1 + b ^2 \Ccal^{-1}(2R_{\rm sym}) \rb.
}
A necessary condition for $P_{11}$ to be positive is for parameter $b$ to satisfy
\ea{
b^4 \leq \f 1 {\Ccal^{-1}(2 R_{\rm sym})\Ccal^{-1}(R_{\rm sym})},
\label{eq:condition b no cooperanios 2 active}
}
similarly, by combining \eqref{eq:schemde power efficiency relay link min powers P11}, \eqref{eq:schemde power efficiency relay link min powers P22} and
\eqref{eq:schemde power efficiency relay link min powers P23} we obtain that a necessary condition for $P_{11}$ to be positive is
\ea{
b^2 \leq \f 1 {\Ccal^{-1}(R_{\rm sym})(1+\Ccal^{-1}(R_{\rm sym}))}.
}
It can be verified that $P_{22}$ and $P_{23}$ can always be determined when $P_{11}$ is feasible.
%
%
This consideration is not surprising: since RX~1 and RX~3 treat the interference as noise, large rates can be achieved only when the level of the interference, which is controlled by $b$ in both cases, is small.
For $b$ to be small it is necessary for $R_{\rm sym}$ to be small as prescribed by \eqref{eq:condition b no cooperanios 2 active}.
Other cases of parameter values can be similarly analyzed with similar conclusions: this scheme employs no cooperation among the relay and it is thus power efficient
on the relay link but does not achieve arbitrarily large rates as it is interference limited.

\subsection{Non-cooperative scheme with two active relays and interference decoding}
\label{sec:Non-cooperative scheme with two active relays and interference decoding}

The achievable scheme in Sec. \ref{sec:Non-cooperative scheme with two active relays} can be improved upon by allowing some receivers to decode an interfering codeword.
Consider again \eqref{eq:schemde power efficiency relay link R2}: in this case the rate $R_2$ is limited by the power of the codeword for $W_1$:
a possible solution for this problem is to allow interference decoding at RX~2.

To show the advantages provided by interference decoding consider the case where $a>1>b$: in this case a well performing achievable scheme that does not
 consider cooperation is the scheme where RN~2 transmits $W_1$ and RN~1 transmits $W_3$.
  $W_2$ can be sent to RN~1 and, since $a>b$, superimposed over the codeword for $W_3$
(in this, the scheme considered here differs from the scheme in  Sec. \ref{sec:Non-cooperative scheme with two active relays}, in which $W_3$ is superimposed over $W_2$. ).
Since $a>b$, the transmission from RN~2 toward RX~1 interfere with the transmission from RN~1 to RX~2: in this case it is convenient to have RX~2 decode $W_1$.
With this choice, while RX~1 and RX~3 only decode their intended message, RX~2 decodes all the messages.
%


This corresponds to the  CGRAS number~2 in \ref{fig:Schemes} (same as the CGRAS number~3 in Fig. \ref{fig:rate01}) with achievable rate region:
\eas{
R_1 & \leq \Ccal \lb \f{P_{11}}{1 + b^2 (P_{22} + P_{23}) } \rb
\label{eq:non coop int decoding R1} \\
R_1 + R_2 + R_3 & \leq \Ccal \lb a^2(P_{11}+ P_{22} + P_{23})\rb
\label{eq:non coop int decoding R1 R2 R3} \\
R_2 + R_3 & \leq \Ccal \lb a^2( P_{22} + P_{23})\rb
\label{eq:non coop int decoding R2 R3} \\
R_1+ R_2 & \leq \Ccal \lb a^2(P_{11}+ P_{22} )\rb
\label{eq:non coop int decoding R1 + R2}\\
R_2 & \leq \Ccal \lb  a^2 P_{22} \rb
\\
R_1 & \leq \Ccal \lb a^2 P_{11}     \rb
 \\
R_3 & \leq \Ccal \lb \frac{P_{23}}{1 + b^2 P_{11} + P_{22}} \rb ,
\label{eq:non coop int decoding R3}
}{{ \label{eq:non coop int decoding}}}
with the assignment
\eas{
X_1 & = \sqrt{P_{11}} U_{1 \sgoes 12}(W_1)\\
X_2 & = \sqrt{P_{22}} U_{2 \sgoes 2}(W_2)  + \sqrt{P_{23}}  U_{2 \sgoes 23}(W_3),
}
which has a total power consumption at the access link of
\ea{
P^{\rm RN}_1+P^{\rm RN}_2= P_{11} + P_{21} + P_{23}.
}

In this case the rate bounds \eqref{eq:non coop int decoding R1} and \eqref{eq:non coop int decoding R3} are still interference
limited as in the scheme in \eqref{eq:schemde power efficiency relay link} but the remaining bounds are not.
On the other hand though, the additional bounds of \eqref{eq:non coop int decoding R1 R2 R3}, \eqref{eq:non coop int decoding R2 R3} and \eqref{eq:non coop int decoding R1 + R2} are not present in \eqref{eq:schemde power efficiency relay link}.
By increasing the number of messages decoded at a terminal, we impose additional constraints on the achievable rate region, since the correct decoding of the interfering codewords has to be possible.
The additional rate constrains might be loose in a subset of the parameter region and thus interference decoding can provide a rate advantage in some class of channels.
Since \eqref{eq:non coop int decoding R1} and \eqref{eq:non coop int decoding R3} are still interference limited,
it is possible that no power allocation exists to achieve a given target rate vector.

\subsection{Partially cooperative scheme}
\label{sec:Partially cooperative scheme}
%
%
We now consider the case it which the relay nodes cooperate in transmitting in transmitting one message. This corresponds to the smallest level of cooperation which corresponds to a base station power consumption of $P^{\rm BS}=2 \cdot 16^{R_{\rm sym}}-2$.

Consider the scenario for $b>1>a$: since $a$ is small it is necessary to for the two RNs to cooperate in transmitting $W_2$ and take advantage of the combining gain.
Since $b>a$ it is more convenient to send $W_3$ through RN~1 and $W_1$ through RN~2 and to superimpose both codewords over the codeword for $W_2$.

The resulting CGRAS is CGRAS number~3 in Fig. \ref{fig:Schemes} (same as CGRAS number~15 Fig. \ref{fig:cgras_legend}) with achievable rate region:
\eas{
R_1+R_2 & \leq  \Ccal \lb \f{ b^2 P_{21} + (b+1)^2 P_2}{1  + P_{13}} \rb \\
    R_1 & \leq  \Ccal \lb \f{ b^2 P_{21} }{1  + P_{13}} \rb\\
R_3+R_2 & \leq  \Ccal \lb \f{ b^2 P_{13} + (b+1)^2 P_2}{1  + P_{21}} \rb \\
    R_3 & \leq  \Ccal \lb \f{ b^2 P_{13} }{1  + P_{21}} \rb  \\
R_2 & \leq \Ccal \lb \f{ 4 a^2 P_2}{1 + a^2 P_{13}+ a^2 P_{21}} \rb,
}{\label{eq:Partially cooperative scheme gaussian region}}
with the assignment
\eas{
X_1 &= \sqrt{P_{2}}U_{1 2 \sgoes 123}(W_2) + \sqrt{P_{13}} U_{1 \sgoes 3}(W_3) \\
X_2 &= \sqrt{P_{2}}U_{1 2 \sgoes 123}(W_2) + \sqrt{P_{21}}  U_{2 \sgoes 1}(W_1),
}
which has a total power consumption at the access link of
\ea{
P^{\rm RN}_1+P^{\rm RN}_2= 2 P_{2} + P_{21} + P_{13}.
}
Note that we have chosen the scaling of $U_2$ in $X_1$ and $X_2$ that provides the maximal ratio combining at RX~2.
Cooperation among the RNs in transmitting $W_2$ provides a combining gain at all the RXs.
For RX~1 and RX~3 this combining gain is $(b+1)^2 P_2$ while for RX~2 is $4 a^2 P_2$, which is the
largest combining gain that can be attained when fixing the total transmission power for $W_2$ at the RNs.

Note that, since $W_2$ is the base codeword, it is also decoded at RX~1 and RX~3 and thus this
combining does not result in further interference for the non-intended decoders.

\subsection{Partially cooperative scheme with interference decoding}
\label{sec:Partially cooperative scheme with interference decoding}

The rate bounds in \eqref{eq:Partially cooperative scheme gaussian region} are all reduced
by the power of the interference experienced at each decoder.
We seek to improve on this scheme by including interference decoding: we do so by allowing
RX~1 and RX~3 to decode every codeword.
%

The resulting CGRAS number~4 in Fig. \ref{fig:Schemes} ( same as CGRAS number~18 in Fig. \ref{fig:cgras_legend}) which achieves the rate region:
\eas{
R_1+R_2+R_3 & \leq  \Ccal \lb b^2 P_{21} + (b+1)^2 P_2 +P_{13} \rb \\
    R_1+R_3 & \leq  \Ccal \lb b^2 P_{21}+ P_{13} \rb\\
R_1+R_2 & \leq  \Ccal \lb b^2 P_{21} + (b+1)^2 P_2\rb \\
    R_1 & \leq  \Ccal \lb b^2 P_{21}  \rb\\
    R_3 & \leq  \Ccal \lb P_{13} \rb\\
R_1+R_2+R_3 & \leq  \Ccal \lb b^2 P_{13} + (b+1)^2 P_2 +P_{21} \rb \\
    R_1+R_3 & \leq  \Ccal \lb b^2 P_{13} + P_{21} \rb  \\
R_3+R_2 & \leq  \Ccal \lb b^2 P_{13} + (b+1)^2 P_2\rb \\
    R_3 & \leq  \Ccal \lb b^2 P_{13} \rb  \\
    R_1 & \leq  \Ccal \lb P_{21} \rb  \\
R_2 & \leq \Ccal \lb \f{ 4 a^2 P_2}{1 + a^2 P_{13}+ a^2 P_{21}} \rb,
}{\label{eq:Partially cooperative scheme with int dec gaussian region}}
with the assignment
\eas{
X_1 &= \sqrt{P_{2}}U_{12 \sgoes 123}(W_2) + \sqrt{P_{13 }} U_{1 \sgoes 13}(W_3) \\
X_2 &= \sqrt{P_{2}}U_{12 \sgoes 123}(W_2) + \sqrt{P_{21 }} U_{2 \sgoes 13}(W_1).
}
%
As for the scheme is Sec. \ref{sec:Non-cooperative scheme with two active relays and interference decoding}, interference decoding improves the SINR at RX~1 and RX~3 but it does impose additional constraint on the rate of the interfering codewords being decoded.

In particular,  for $b>1>a$, the rate of $R_1$ ($R_3$) is reduced to $\Ccal(P_{21})$ ($\Ccal(P_{13})$) from $\Ccal(b^2 P_{21})$ ($\Ccal(b^2 P_{13})$)
since correct decoding must take place at RN~3 (RN~1)  as well.

\section{Simulation Setup}
\label{sec:Numerical Evaluations}

%
%

Through numerical simulation, we can consider a large number of schemes and select the scheme corresponding to the smallest power consumption.
The number of achievable transmission strategies which can be generated using \cite[Th. IV.2.]{RiniEnergyPartI13} is on the order of hundreds
without rate-splitting and thousands with rate-splitting.
The power optimization for each scheme can be efficiently performed using linear optimization tools under some mild restrictions
on the power allocation for the cooperative schemes.

Indeed the number of schemes considered and the low computational complexity make it possible to identify a very detailed transmission
 strategy with reasonable resources.

%

%
In the following we consider the following numerical optimization problems:
\begin{itemize}
  \item {\bf No rate-splitting Scenario:}  When no rate-splitting is considered, the number of possible achievable rate regions is limited and it is possible to graphically present the results of the energy efficiency optimization for varying $a$ and $b$ parameters.
      We also connect the numerical results with the schemes in Sec. \ref{Sec:Considerations on a Simple Example} to provide further intuitions.
  \item {\bf Rate-Splitting Scenario:}  when considering both superposition coding and rate-splitting there are simply too many schemes to be efficiently presented graphically.
   In this case we group the schemes according to their features in terms of the cognition level, number of superposition coding steps and
    amount of  interference decoding.
    This description is less detailed than the previous scenario, but still insightful on a higher level perspective as in Sec. \ref{sec:Problem Description and an Illustrative Example}.
  \item {\bf Comparison between Inner and Outer Bounds:} we compare the results of the numerical optimization with the lower bound
  on the energy consumption in Sec. \ref{sec:Lower Bounds to the Energy Efficiency}  and the case of relay selection and uncoordinated transmissions.
       For this scenario we consider the throughput comparison among different strategies for fixed channel parameters.
\end{itemize}

\subsection{No Rate-Splitting Scenario}
\label{sec:No rate-splitting Scenario}

We begin by considering the achievable schemes that can by generated from \cite[Th. IV.2.]{RiniEnergyPartI13} which do not involve rate-splitting.
By restricting our attention to this class of schemes, the optimization results are limited to eighteen  schemes (after accounting for symmetry).
Given the limited set of schemes, we can provide various intuitions on the important features of the optimal solution for varying $a$ and $b$.

\begin{figure}[h]
\centering
\includegraphics[width=1\textwidth]{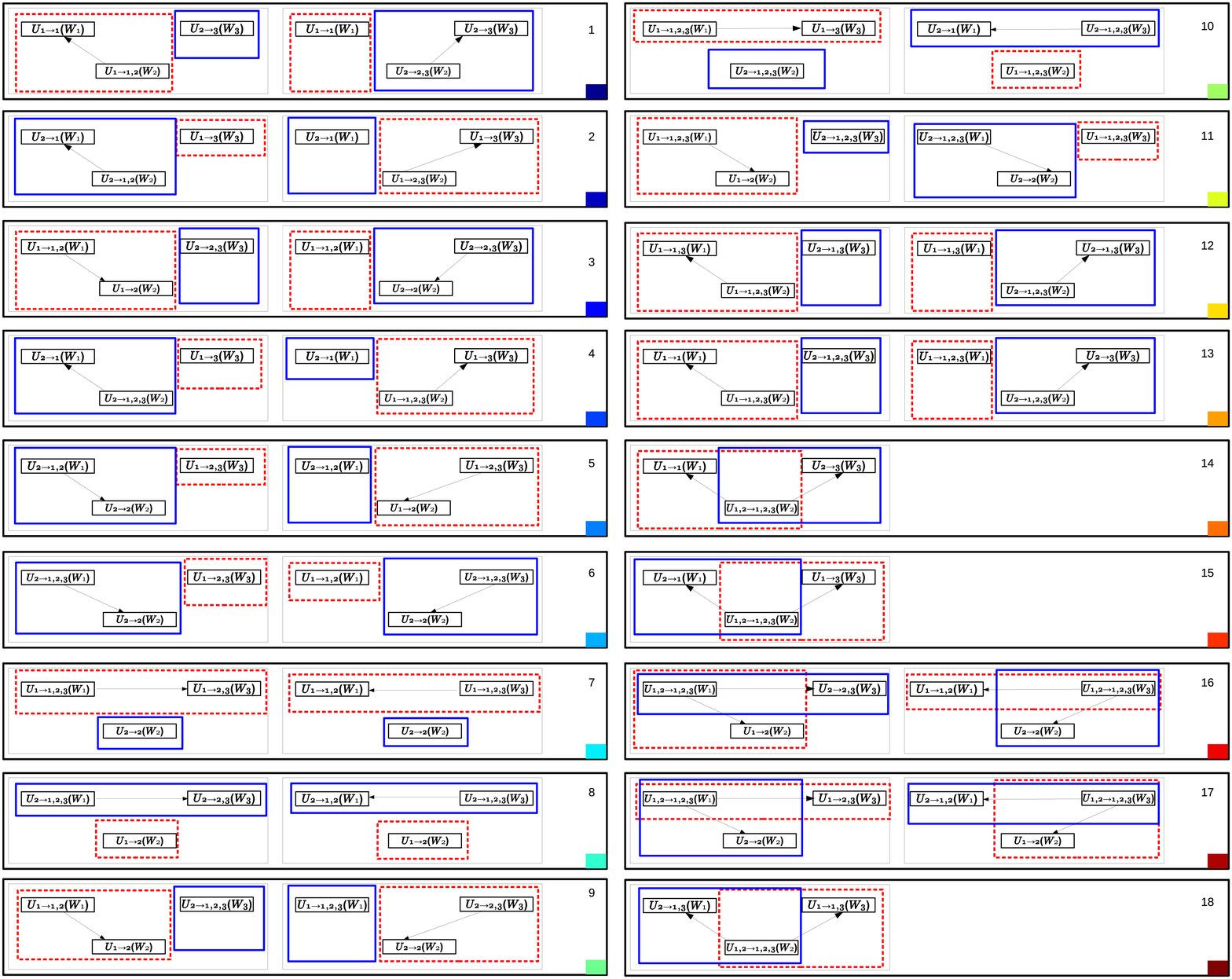}
\vspace{- 7.4 cm}
\caption{
A graphical representation of the best performing CGRASs for the scenario in Sec. \ref{sec:No rate-splitting Scenario}.
}
\label{fig:cgras_legend}
\end{figure}

The solution of the optimal energy efficiency problem are represented using the corresponding CGRAS as in Fig. \ref{fig:cgras_legend}.
%
%
Each CGRAS is contained in a different partition of the figure with an associated number and color:
The schemes are arranged by increasing level cooperation and increasing interference decoding:

\begin{itemize}
  \item {\bf blue tones, numbers 1--8:}   no cooperation and minimum or no interference decoding schemes,
  \item {\bf green and yellow tones, numbers 9--13:} no cooperation and interference decoding schemes,
  \item {\bf red tones, numbers  14--18:}  partial cooperation schemes with some interference decoding.
\end{itemize}

In each box, two rectangles represents the two RNs: the rectangle with a red, dotted edge is associated with RN~1 while the one with a blue, solid edge to RN~2.
Each rectangle contains the CGRAS nodes known at the associated RN. If a codeword is known at both, in is placed in the intersection of the two rectangles.
Each node of the CGRAS is represented by a box labeled as $U_{\iv \sgoes \jv} (W_z)$.
As in \cite[Def. 1]{RiniEnergyPartI13},  $U_{\iv \sgoes \jv}$ is the RV associated with  the codeword transmitted from the set of transmitter $\iv$ toward of
the set of receiver $\jv$ while $W_z$ indicates that this codeword embeds the message $W_z$ as in Sec. \ref{Sec:Considerations on a Simple Example}.
Superposition coding is indicated as a directed arrow, from the base codewords to the top one.
Due to the symmetry in the channel model and in the target rate vector, more than one scheme achieves the same achievable rate region.
In particular, given any scheme, swapping the labeling of  RN~1 and RN~3 and swapping 2 to 1 in $\iv$ and 1 and 3 in $\jv$  for all
the codewords $U_{\iv \sgoes \jv}$ will result in a scheme with the same achievable rate region.
%
%
We refer to these schemes as equivalent and equivalent schemes are placed in the same partition of the figure.

%
%

The optimization is performed as follows: for each possible message allocation among the RNs, all possible combinations of superposition coding and interference decoding are considered.
With each such choice, a CGRAS is obtained along with the corresponding achievable rate region through \cite[Th. IV.2.]{RiniEnergyPartI13}.
The achievable rates are evaluated for the choice of auxiliary RV in \cite[Lem. IV.3.]{RiniEnergyPartI13}, additionally the scaling of the cooperative
codewords is chosen so as to provide the largest combining gain at the intended receiver.
With this choice of distribution of $U_{\iv \sgoes \jv}$, the  power consumption is a linear function of the matrix $A$ in \cite[eq. (27)]{RiniEnergyPartI13} and
therefore the minimum power consumption on the access link for which the target symmetric rate is achievable can be determined using standard linear programming algorithms.
The total power consumption is obtained by adding the power consumption at the access link with the power consumption on the relay link which is obtained through
\cite[Th. V.1.]{RiniEnergyPartI13} for each message allocation at the RNs.
This optimization is repeated over multiple channel parameters $(a,b)$.
For each channel, the optimal schemes are selected with a  5\% tolerance and when more than one scheme is within the prescribed tolerance, we choose the scheme
which is optimal for values of similar of $(a,b)$.
In the following, we consider the four symmetric rates: $R_{\rm sym} \in [0.1 , 0.5 , 1 , 2]$ and the range of channel parameters $a \times b= [0 \ldots 2]^2$.

\begin{figure}[ht]
\centering
\subfigure[$R_{\rm sym} = 0.1$ ]{
\includegraphics[width=0.47\textwidth]{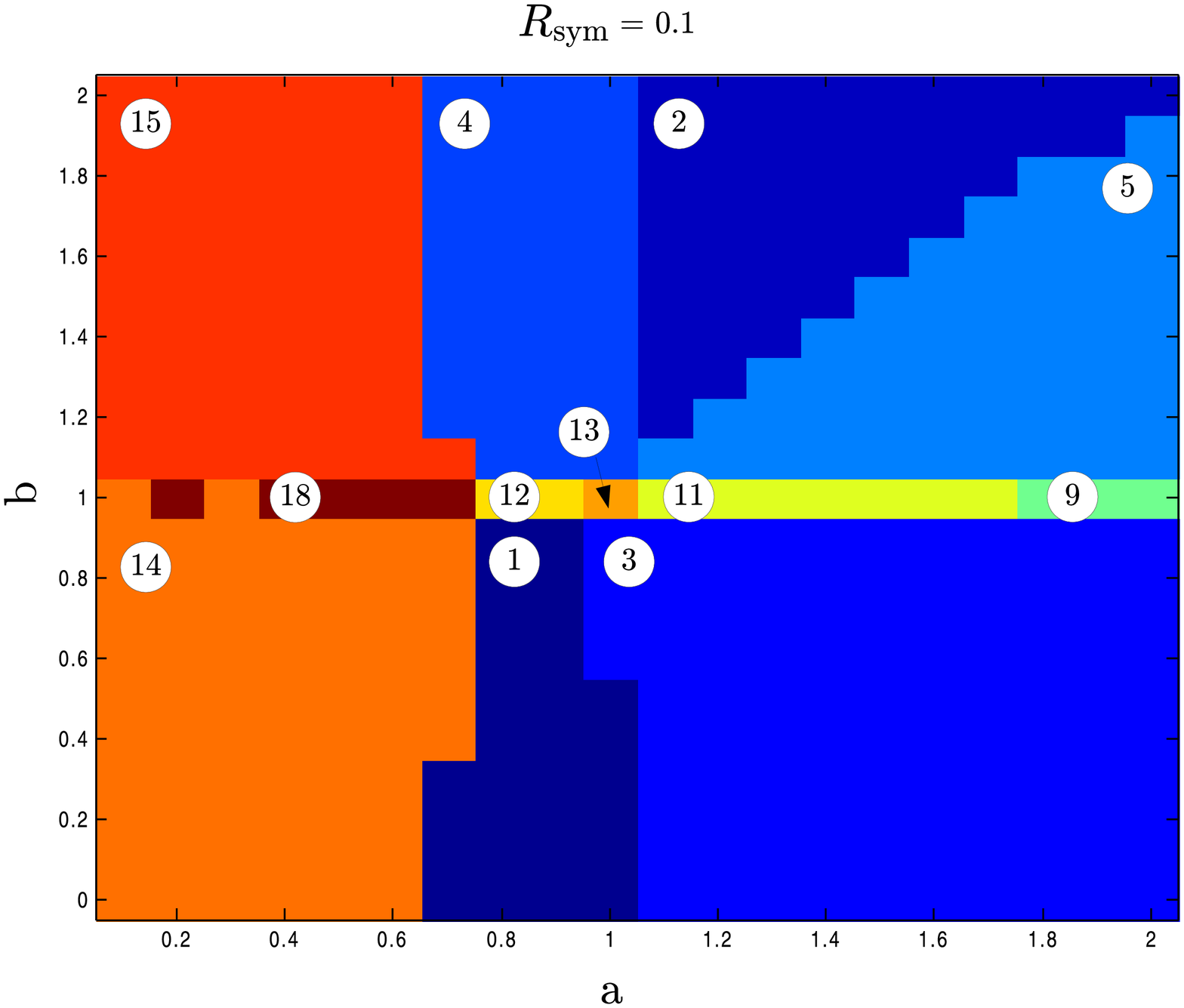}
\label{fig:rate01}
}
\subfigure[$R_{\rm sym} = 0.5$ ]{
\includegraphics[width=0.47\textwidth]{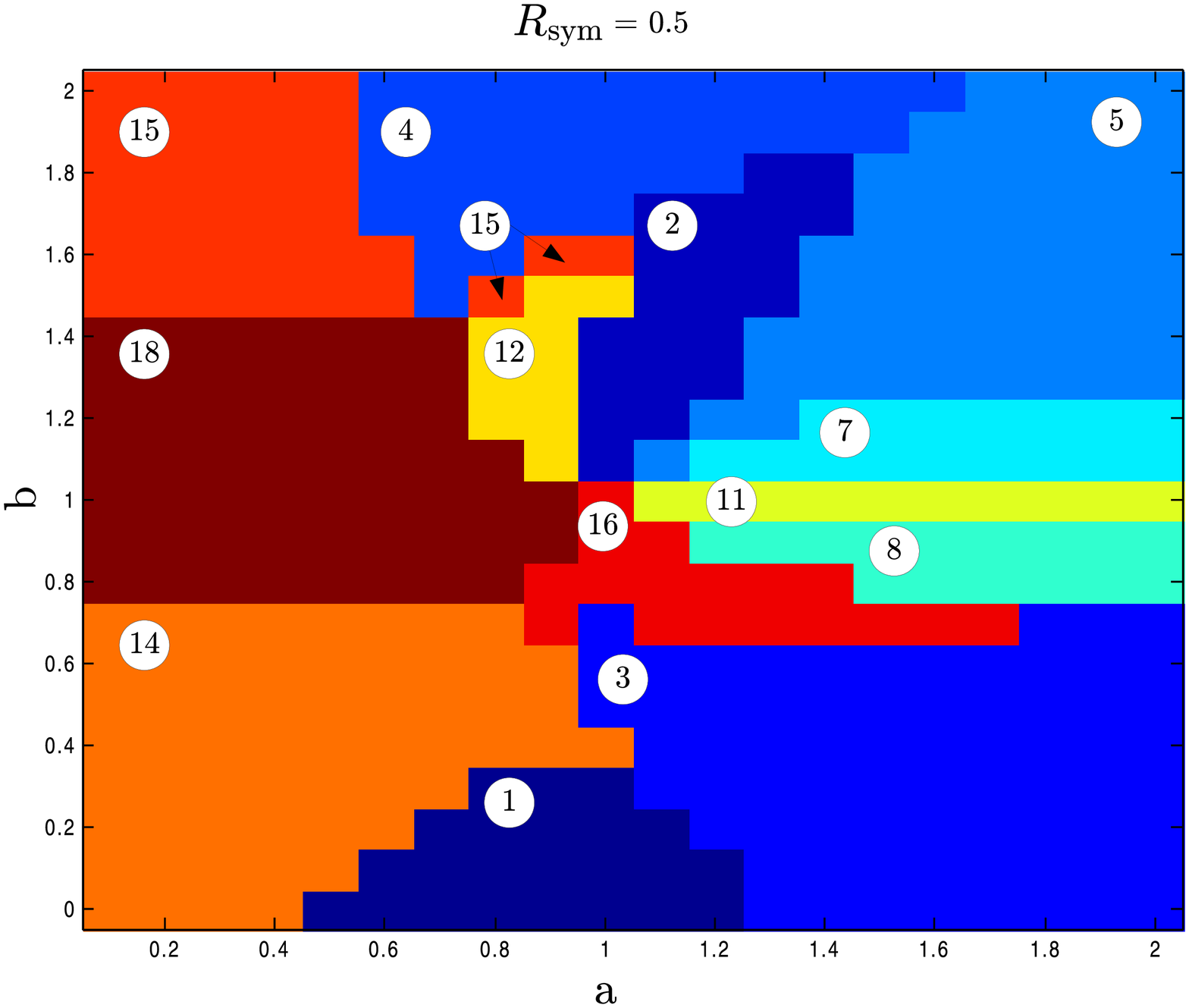}
\label{fig:rate05}
}
\caption{Numerical simulations for the best CGRAS performance for $R_{\rm sym} = 0.1$ and $R_{\rm sym} = 0.5$  for $a \times b \in [0,2]^2$. }
\label{fig:Numerical simulations I}
\end{figure}

\medskip
%
%
Let's start by considering the case of $R_{\rm sym}=0.1$: the optimization results are provided in Fig. \ref{fig:rate01}.
Since the achievable rate is relatively low, we expect the scheme of Sec. \ref{sec:Non-cooperative scheme with two active relays} to be both feasible and efficient
in achieving $R_{\rm sym}$. Indeed scheme~1 and~2, which  correspond to the encoding choice of Sec. \ref{sec:Non-cooperative scheme with two active relays},
are optimal around the region $b>a>1$ and the region $b<1=a$.
In these two regions the level of the interference at each receiver is limited and treating the interference as noise is optimal.
Also, there is no need for cooperation as the advantages provided by the combining gains do no outweigh the additional costs of cooperation on the relay link.
A detailed explanation of scheme~1 is provided in Sec. \ref{sec:Non-cooperative scheme with two active relays}.
For the region $b>a>1$, we notice that RN~1/ (RN~2) can communicate to RX~3/ (RX~1) through a link with gain $b$ at high SNR.
RX~2 can be served by either RN~1 or RN~2 also through the link $a$: the presence of simultaneous transmission on this link create excessive interference
to the rate $R_2$ since a large $b$ guarantees that the power of this transmission is low.

For larger values of the parameter $a$, it is more advantageous for RX~2 to decode part of the interference and thus schemes similar to the one in
Sec. \ref{sec:Non-cooperative scheme with two active relays and interference decoding} perform better.
Indeed schemes~3 and~5, having the same coding strategy of the scheme in Sec. \ref{sec:Non-cooperative scheme with two active relays and interference decoding},
are optimal around the region $a>1>b$ and $a>b>1$.
The scheme~3 is presented in Sec. \ref{sec:Non-cooperative scheme with two active relays and interference decoding}, scheme~5
is a simple variation of scheme~3 where RN~1 (RN~2) serves RX~3 (RX~1).

When $a$ is small, then coherent combining in necessary to attain $R_2=R_{\rm sym}$ and the RNs need to cooperate in transmitting message RX~2.
The remaining two messages are superimposed over the codeword for $W_2$ and we thus obtain the cooperative scheme of Sec. \ref{sec:Partially cooperative scheme}
and of scheme~14 and scheme 15. Note that the value of $b$ only determines which RN serves which RX and no interference decoding is required at RX~1 and RX~2 apart for the case $b \approx 1$.

\medskip
Let's consider now the case $R_{\rm sym}=0.5$ presented in Fig. \ref{fig:rate05}.
Since the symmetric rate is higher that the case depicted in Fig. \ref{fig:rate01}, it no longer optimal to treat the interference as noise and thus the region in which scheme~1 and scheme~2 are optimal are
smaller than in Fig. \ref{fig:rate01}. Schemes involving various possible interference decoding, such as scheme~4, 7, 8 and 12 emerge as optimal.
Also Scheme~18, which is detailed in Sec. \ref{sec:Partially cooperative scheme with interference decoding} and which involves cooperation and interference decoding, becomes optimal in a larger region around $b \approx 1$.

\begin{figure}[ht]
\centering
\subfigure[$R_{\rm sym} = 1$.]{
\includegraphics[width=0.47\textwidth]{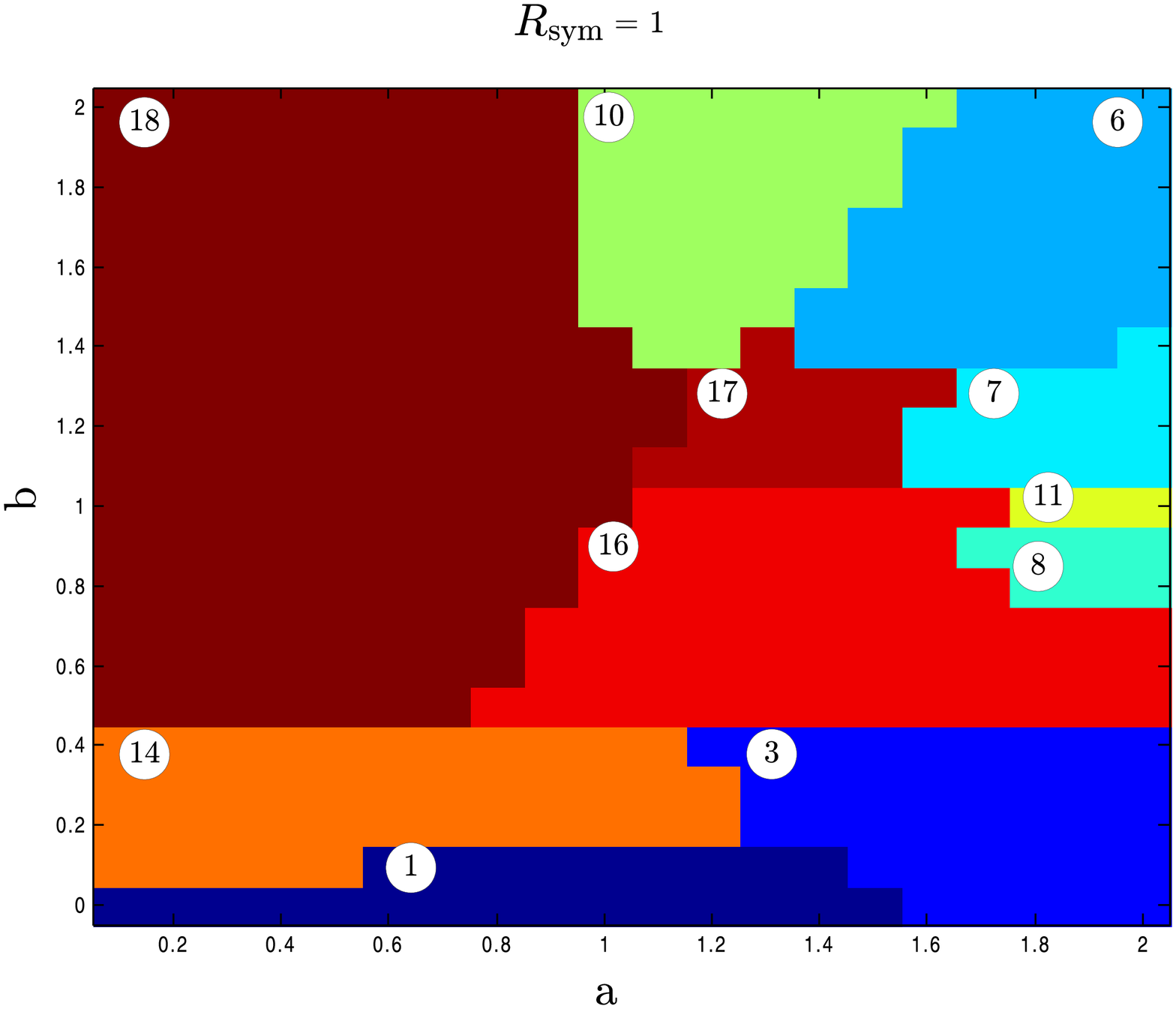}
    \label{fig:rate1}
}
\subfigure[$R_{\rm sym} = 2$.]{
\includegraphics[width=0.47\textwidth]{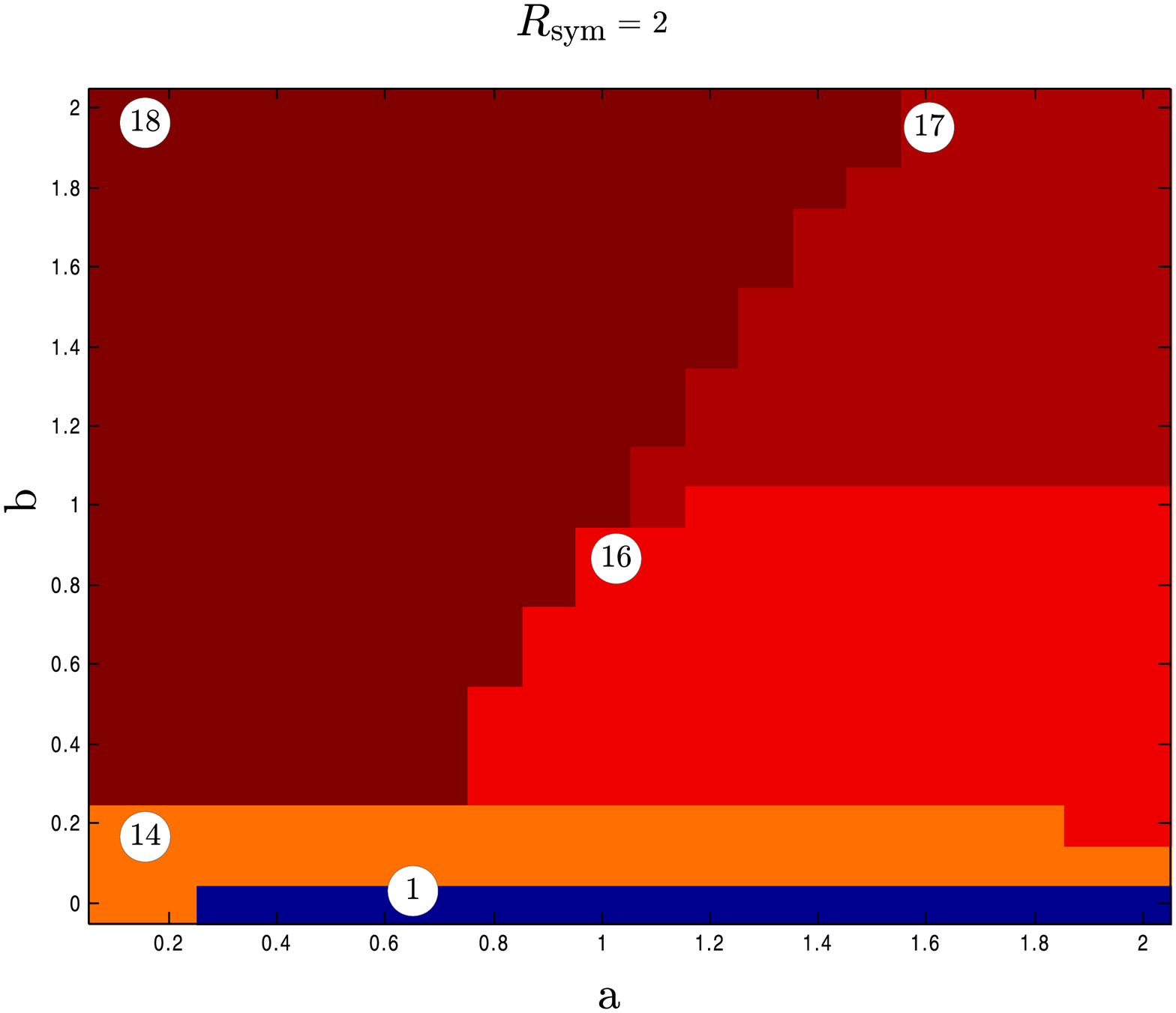}
    \label{fig:rate2}
}
\caption{Numerical simulations for the best CGRAS performance for $R_{\rm sym} = 1$ and $R_{\rm sym} = 2$ for $a \times b \in [0,2]^2$.}
\label{fig:Numerical simulations II}
\end{figure}

As the rate $R_{\rm sym}$ is further increased to $R_{\rm sym}=1$ and $R_{\rm sym}=2$ in Fig. \ref{fig:rate1} and \ref{fig:rate2} we see that  cooperation becomes more
and more advantageous as it allows the RNs to apply superposition coding in order to manage the interference from simultaneous transmissions.
When applying superposition coding,  the receiver of the top codeword also decodes the bottom codewords. In doing so, this receiver also performs interference decoding
and can thus partially remove the effect of the interference.
Although costly on the access link, cooperation makes it possible to perform two superposition coding steps instead of one, and this is the key of the efficiency
of schemes~14, 16, 17 and 18.

%
%
%
%

%
%

\bigskip

\begin{figure}[ht]
\centering
\subfigure[Overall Power Consumption for $R_{\rm sym} = 0.1$.]{
\includegraphics[width=0.45\textwidth]{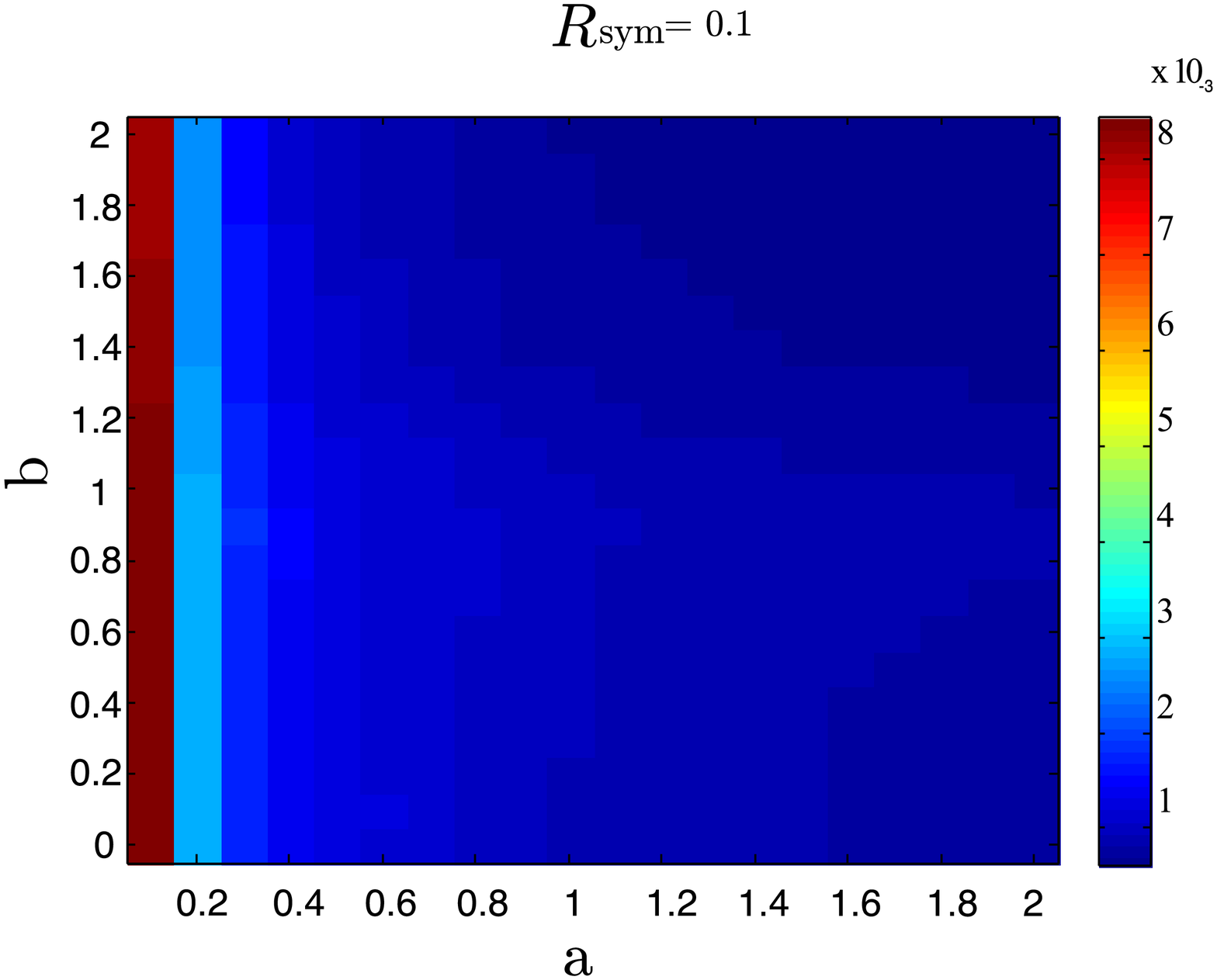}
\label{fig:power01}
}
\subfigure[Overall Power Consumption for $R_{\rm sym} = .5$.]{
\includegraphics[width=0.45\textwidth]{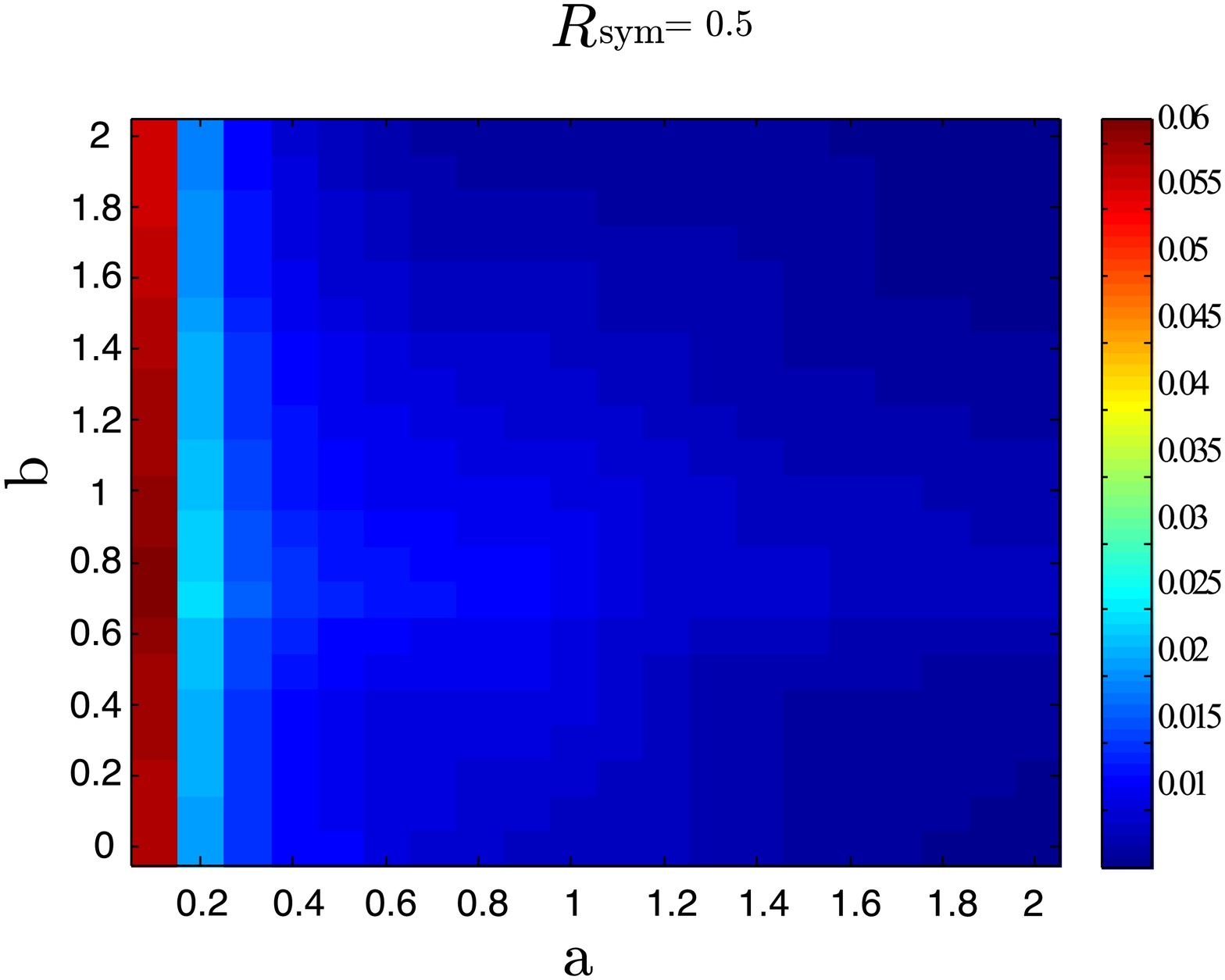}
\label{fig:power05}
}
\subfigure[Overall Power Consumption for $R_{\rm sym} = 1$.]{
 \includegraphics[width=0.45\textwidth]{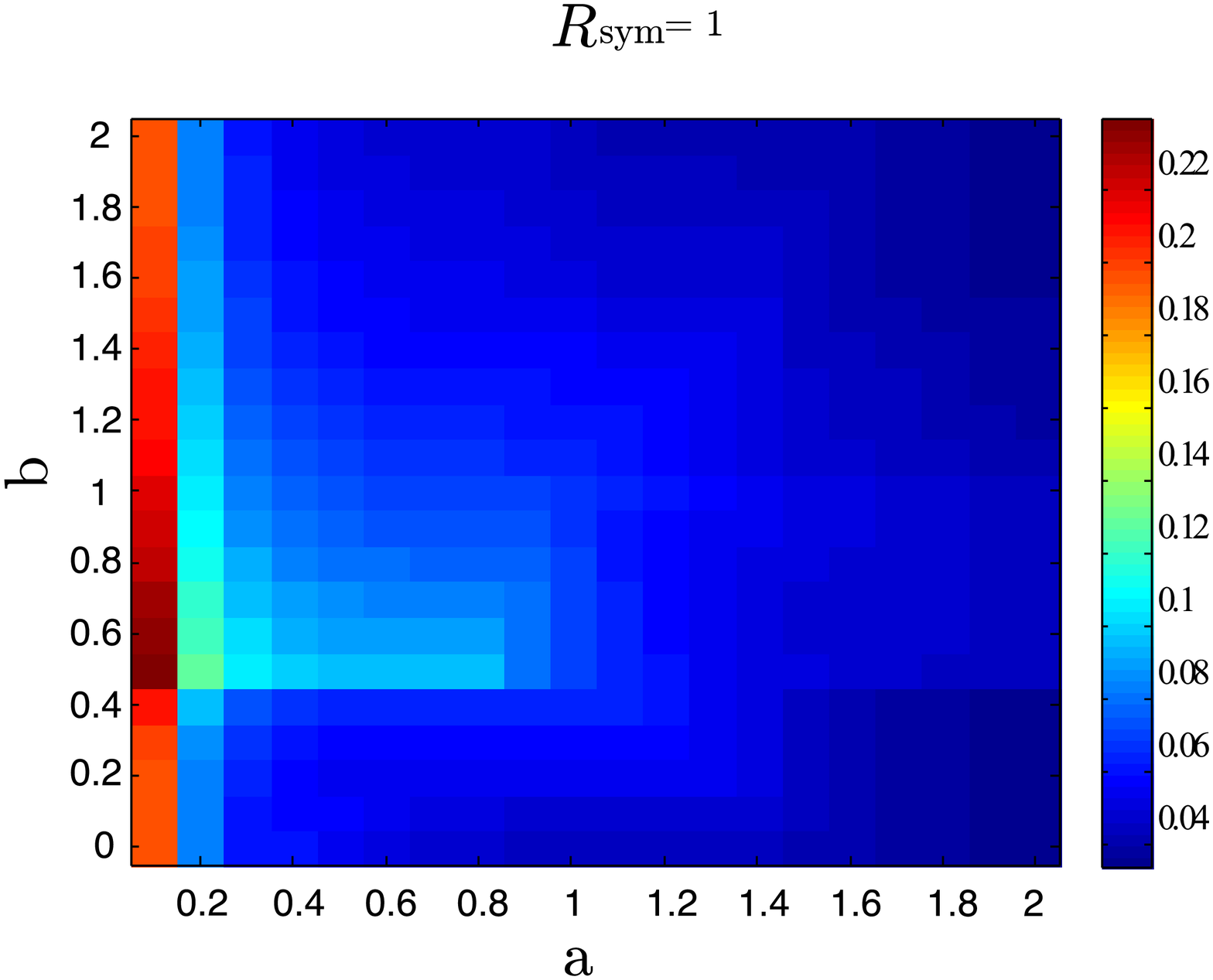}
\label{fig:power1}
}
\subfigure[Overall Power Consumption for $R_{\rm sym} = 2$.]{
\includegraphics[width=0.45\textwidth]{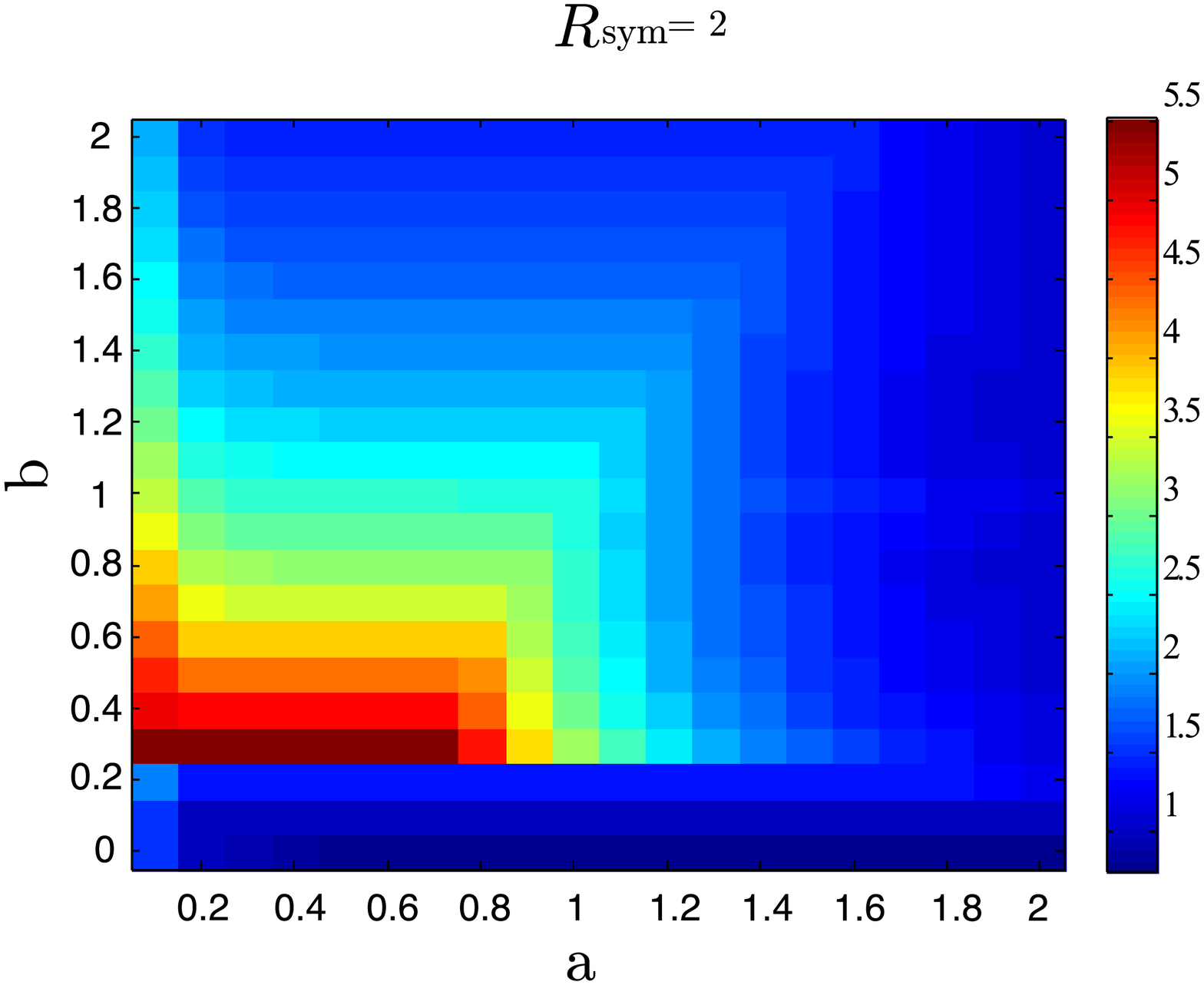}
\label{fig:power2}
}
\caption{Power efficiency for varying symmetric rate $R_{\rm sym}$ and $a \times b \in [0,2]^2$.
}
\label{fig:subfigureExample}
\end{figure}

Fig. \ref{fig:subfigureExample} plots the overall power consumption for the optimization results in Fig. \ref{fig:rate01}, \ref{fig:rate05}, \ref{fig:rate1} and \ref{fig:rate2}.
In all cases the largest power consumption is attained at small values of $a$: since both RNs have a link of gain $a$ toward RX~2, it is not possible to obtain
a better SNR toward this node. Since the communication toward RX~2 is arduous for both receiver in this regime, cooperation is important to provide combining
gains which indeed greatly reduces the power consumption form moderate values of $a$.
Another regime with high power consumption for large symmetric rates is for small $a$ and around $ 0.2 < b < 0.8$: by referring to the optimization results we see that this is a transition point between scheme~14 and scheme~18.
In both schemes the two relays cooperate in transmitting the codeword for $W_2$ and superimpose the messages for RX~1 and RX~3 over this codeword.
The two schemes differ in that scheme~14 does not employ interference decoding at RX~1 and RX~3 while scheme~18 does.
Since $b$ is relatively small, the power at which the codewords for RX~3 is received at RX~1 is not sufficiently strong to be decoded and yet too strong to be
treated as noise.  The same occurs at RX~3 for the interfering codeword for RX~1.
The solution to this impasse is for both decoders to increase their transmitted power, so as to eventually increase the received SNR.
From a high level perspective, and given the symmetry in the channel and the target rate, the SNR at both RX~1 and RX~3 scales as $P/(1+b^2 P)$ when decoding $W_2$
but treating the remaining interference as noise.
Clearly  an increase in $P$ increases the SNR, although very slowly. For this reason the power consumption in this region is relatively high: a high power $P$ is required to attain the SNR which grants reliable communication at rate $R_{\rm sym}$.

\medskip

The power consumption corresponding to the numerical optimization in Fig. \ref{fig:rate01},  \ref{fig:rate05}, \ref{fig:rate1}   and \ref{fig:rate2}
are shown in Fig. \ref{fig:power01}, \ref{fig:power05}, \ref{fig:power1} and \ref{fig:power2} respectively.
From these figures is clear that, as the symmetric rate increases, the largest power consumption corresponds to the region $a \in [0,1]$ and $b \in [0.2 , 1]$:
which corresponds to the region where the numerical optimal scheme changes from 14 to 18.
In scheme 14 the codeword for $W_3$ ($W_1$) is treated as noise RX~1 (RX~3) while in scheme 18 this codeword is decoded.
In this region, therefore, the large power consumption corresponds to the tension between treating the interference as noise or decoding the interference which increases as the symmetric rate increases.
As we shall see, rate-splitting makes it possible to reduce the power consumption in this region by allowing partial interference cancelation.

%
%

\subsection{Rate-Splitting Scenario}

In this section we analyze the effect of rate-splitting on the power consumption.
Since the number of schemes that can be generated using \cite[Th. IV.2.]{RiniEnergyPartI13} in this case is quite high, it is not longer possible to analyze each scheme separately as in Sec. \ref{sec:No rate-splitting Scenario}.
Instead we focus on higher-level features of the transmission schemes and overall power consumption.

\begin{figure}[ht]
\centering
\subfigure[Numerical simulations for the best CGRAS performance considering rate-splitting for $R_{\rm sym} = 2 $ and $a \times b \in (0,2)^2$.]{
\includegraphics[width=.47 \textwidth]{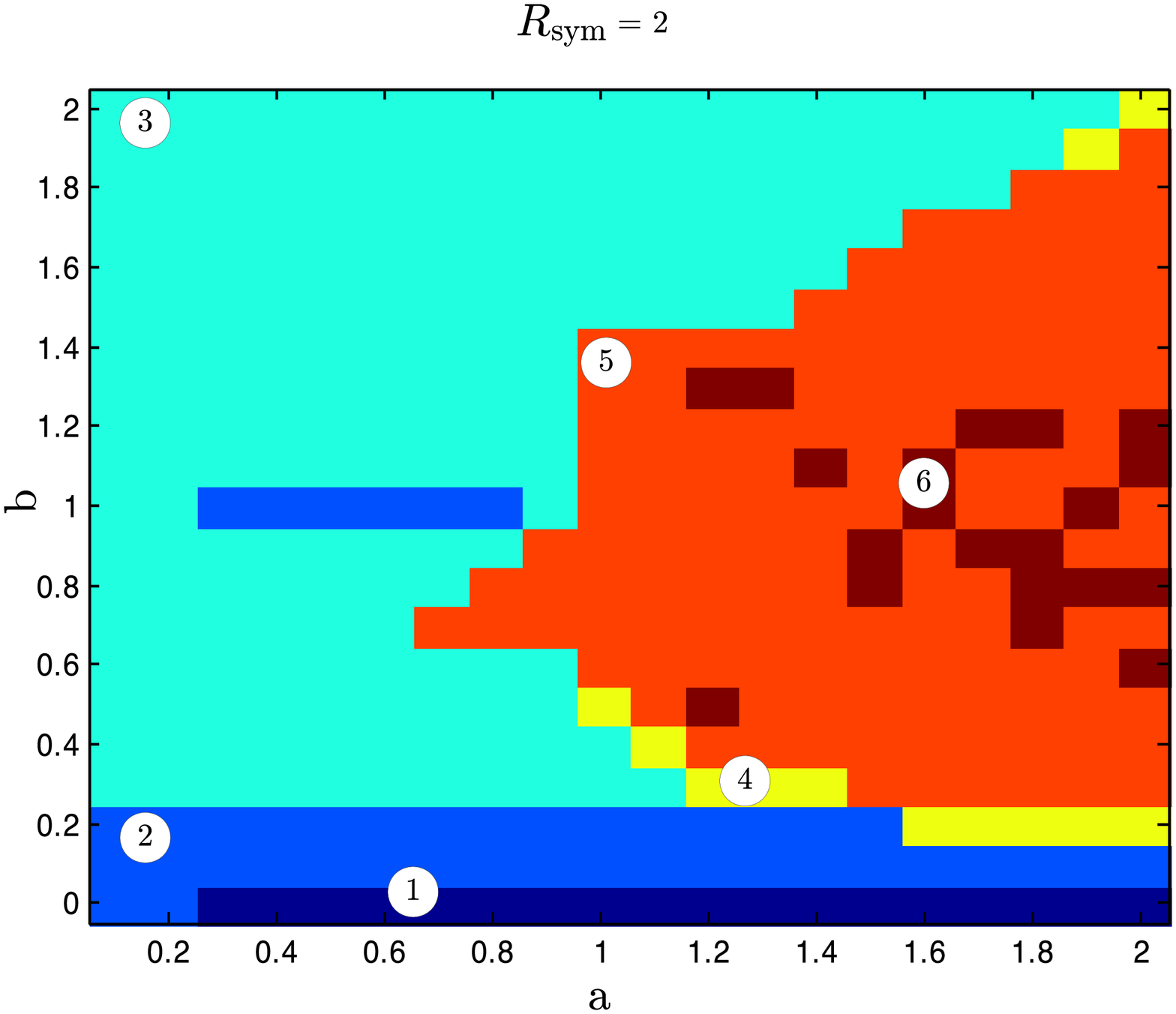}
    \label{fig:rs_rate2}
}
\subfigure[CGRAS Legend for rate-splitting scenario.]{
\includegraphics[width=.47 \textwidth]{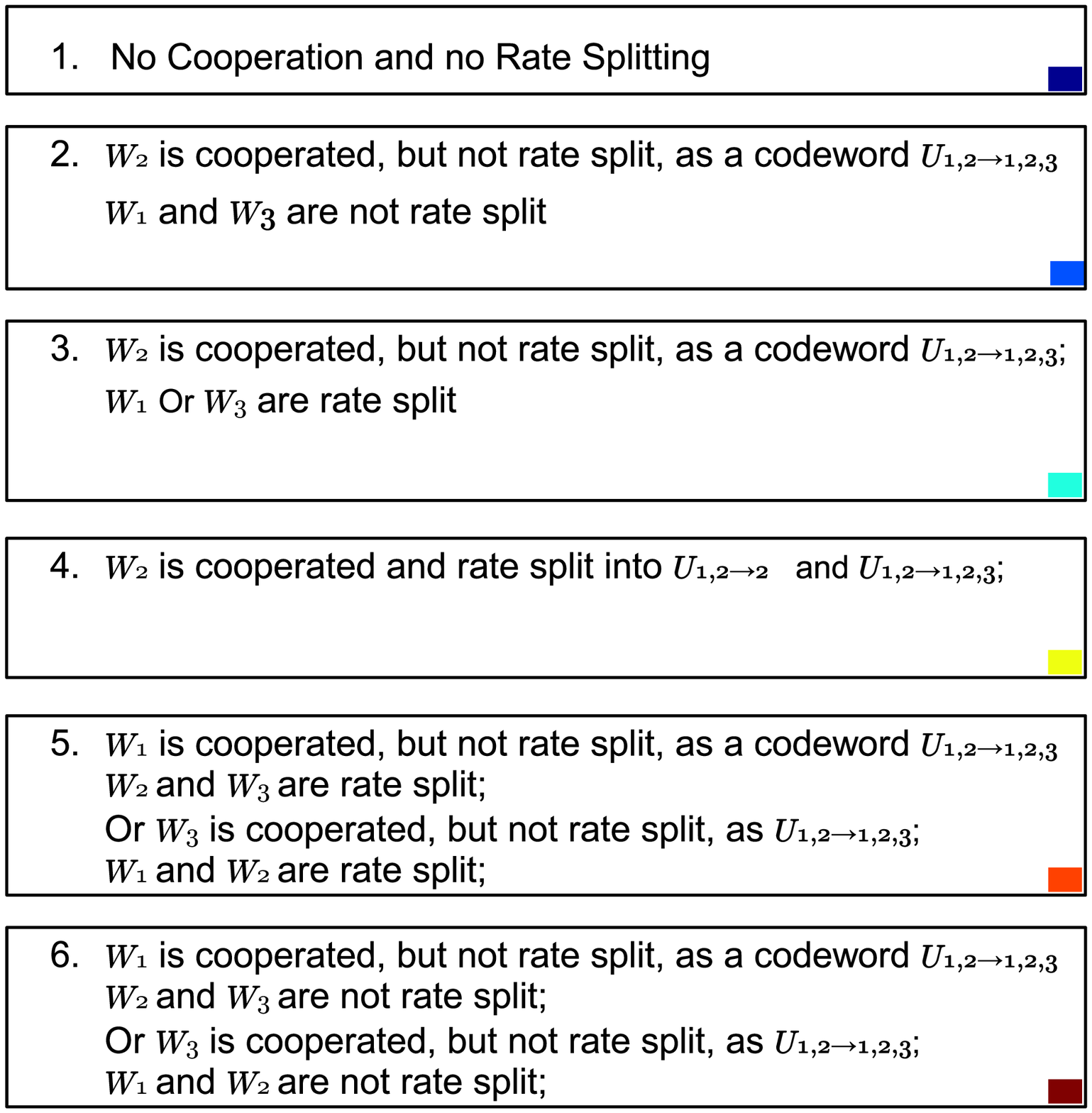}
    \label{fig:rs_legend}
}
\caption{Best CGRAS performance with rate splitting.}
\end{figure}

The solution of the power minimization with rate-splitting for the case $R_{\rm sym}= 2$ is presented in Fig. \ref{fig:rs_rate2} while the legend of this figure is in Fig.  \ref{fig:rs_legend}.
The optimal schemes  are numbered and color coded according to the level of cooperation among the RNs, in particular:

\begin{itemize}
  \item {\bf blue tones, numbers~1 and~2:}   no cooperation with either with or without rate-splitting,
    \item {\bf red tones, numbers~3--5:}  partial cooperation schemes with some interference decoding.
  \item {\bf yellow tones, numbers~6 and~7:} no cooperation and interference decoding schemes.
\end{itemize}

%
%
%

%
%
From Fig. \ref{fig:rs_rate2}  it is clear that rate-splitting always provide an advantage: the schemes which do not employ rate-splitting, scheme 1 and scheme 2, are optimal mostly in the regime where $b$ is close to zero, which corresponds to the case where no interference is created at RX~1 and RX~2.
As $b$ increases, rate-splitting allows for the receiver to decode part of the interfering codeword and strip it from the received signal.
For value of $a<1$, the optimal scheme often involve cooperation over $W_2$: since the link toward RX~2 is relatively weak, the RNs need not cooperate in the transmission of the codeword associated to this message. As the power use to transmit this codeword combines coherently at all the RXs, it is more advantageous for all RXs to decode it, as in scheme 14 and 18 in Fig. \ref{fig:rate2}, and no rate-splitting is necessary.
rate-splitting is instead useful in the transmission of $W_1$ and $W_3$ as it allows a RX to decode part of the interfering codeword.
%

For the case $a>1$, cooperation is no longer necessary to transmit $W_2$  and is instead useful when sending $W_1$ or $W_3$: as for scheme 16 and 17 in Fig. \ref{fig:rate2} the codeword for either message is decoded at all the receivers and acts as a base codeword to transmit the other messages while, again, rate-splitting is useful in allowing for partial cancelation of interfering codewords.

\begin{figure}[ht]
\centering
\subfigure[Power efficiency with rate-splitting for $R_{\rm sym}=2$ and $a \times b \in (0,2)^2$.]{
\includegraphics[width=.47 \textwidth]{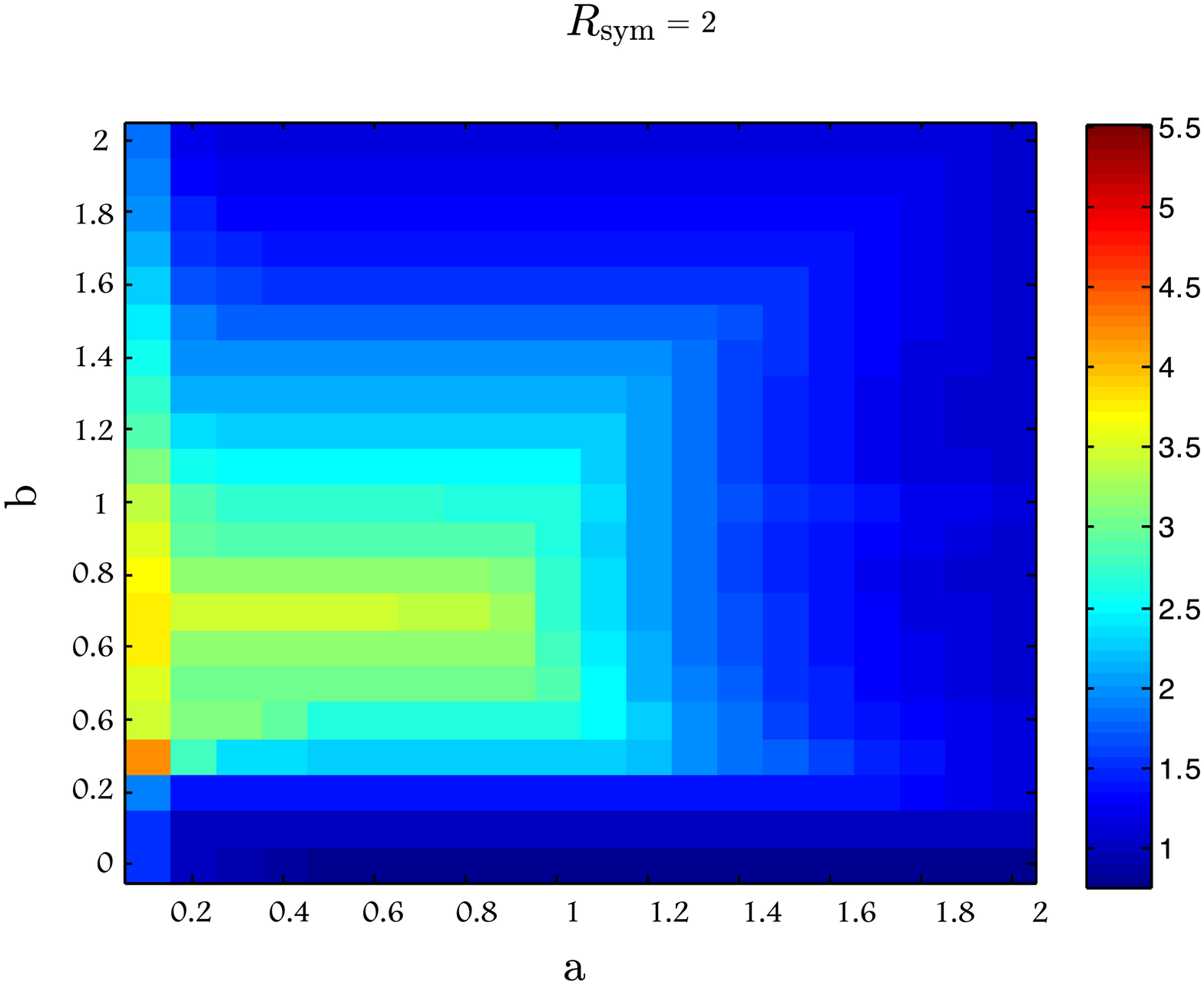}
\label{fig:power2rs}
}
\subfigure[Performance comparison of rate-splitting and no rate-splitting scenarios for $R_{\rm sym} = 2$ and $a \times b \in (0,2)^2$.]{
\includegraphics[width=.47 \textwidth]{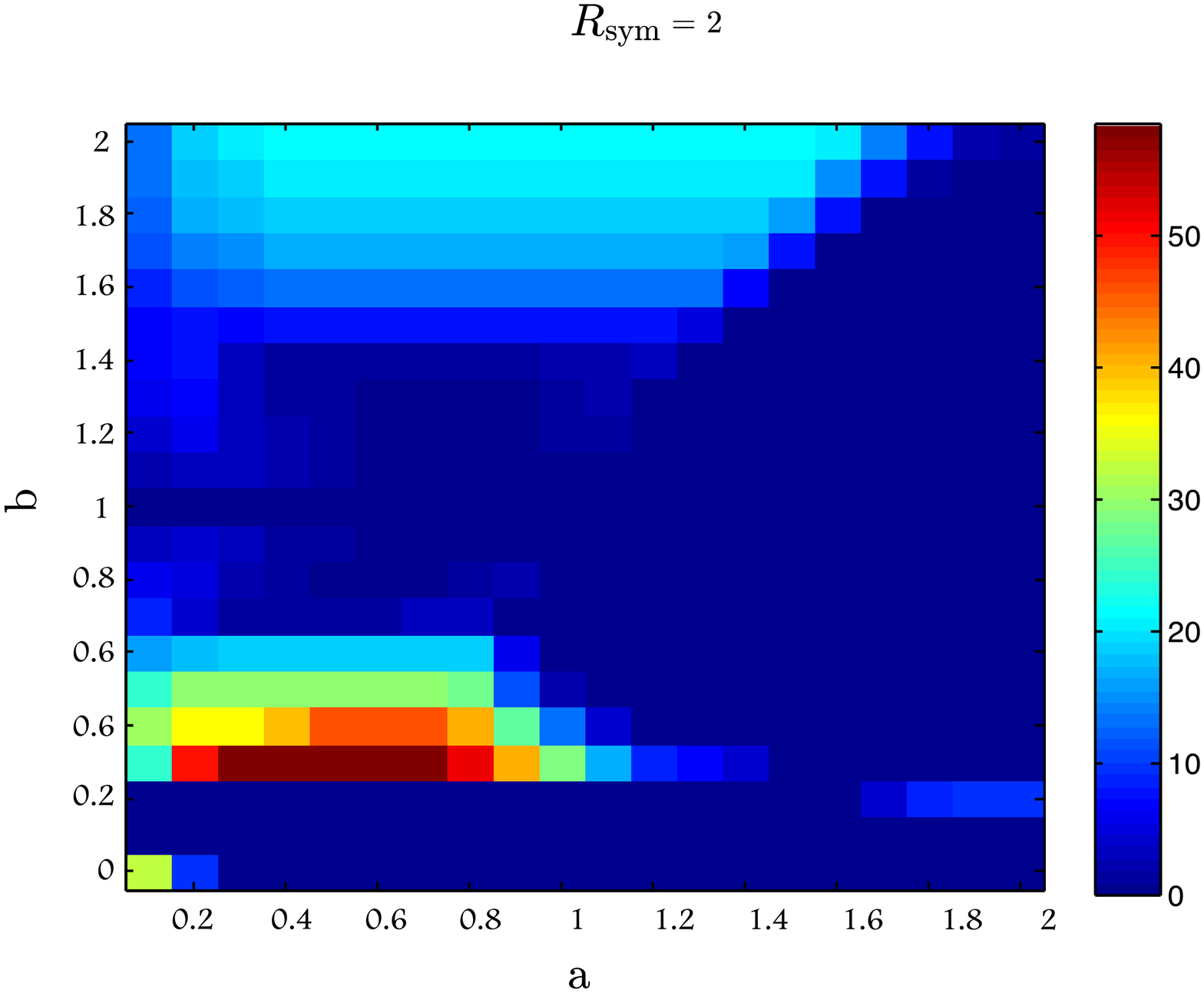}
\label{fig:rs_diff}
}
\caption{Power efficiency with rate-splitting}
\end{figure}

The power consumption associated with the numerical optimization in Fig. \ref{fig:rate2} is presented in Fig. \ref{fig:power2rs} and can be compared to the non rate-splitting  power consumption in Fig. \ref{fig:power2}.
As for Fig. \ref{fig:power2},  the largest power consumption is in the range $a \in [0,1]$ and $b \in [0.2 , 1]$. The difference is the two figures is the regime $a \approx 0.3$: without rate-splitting, this regime has the largest power consumption, whether with rate-splitting  the power consumption in this regime is greatly reduced.

Fig. \ref{fig:rs_diff} plots the power efficiency improvement of rate-splitting case with respect to the non rate-splitting case.
The power improvement are relatively small for the class of channel in which a low power consumption can already be attained without rate-splitting.
In the cases in which the performance of the system are more severely limited by the interference, rate-splitting attains substantial power efficiency gains.

%
%
%
%
%
%
%

\subsection{Comparison between Inner and Outer Bounds}

\begin{figure}[h]
\centering
\includegraphics[width=0.5\textwidth]{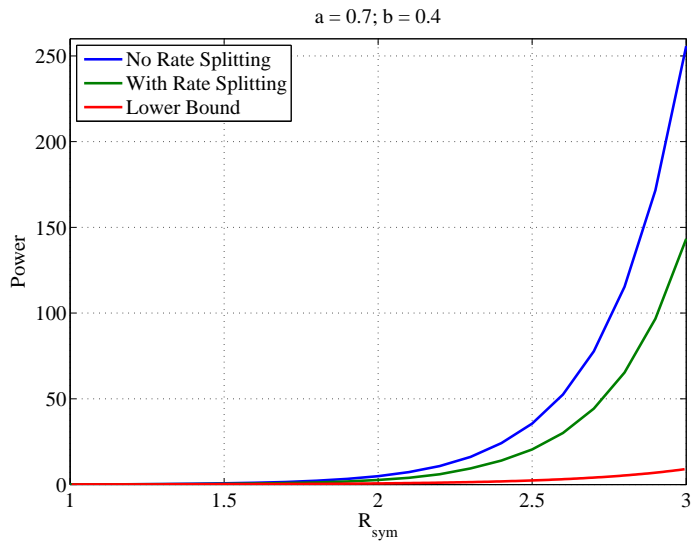}
\vspace{ 0.1 cm}
\caption{
Performance comparison of rate-splitting and no rate-splitting scenarios with outer bound for $R_{\rm sym} \in [1,3]$, $a = 0.7$ and $b  = 0.4$.
}
\label{fig:outerbound}
\end{figure}

In this section we compare the lower bound  in Sec. \ref{sec:Lower Bounds to the Energy Efficiency} with the upper bound in \cite[Sec. IV.B.]{RiniEnergyPartI13}. The upper bound is evaluated through numerical optimization while the lower bound is evaluated through Monte-Carlo simulations.
A plot of the upper and lower bound on the energy consumption for the channel model in Fig. \ref{fig:simsetup_a_b} for $a=b=1$ and increasing symmetric rate $R_{\rm sym}$ is presented in Fig. \ref{fig:outerbound}.
This figure shows that transmission schemes involving rate-splitting provides clear energy saving over the schemes without rate-splitting, especially at high symmetric rates.
Unfortunately, the gap between the best upper bound and the lower bound increases as the symmetric rate increases: the lower bound in Sec. \ref{sec:Lower Bounds to the Energy Efficiency} is know to be loose for the general channel model, so the divergence between upper and lower bounds is expected.

\subsection{Relevance of the Results for Practical Networks}

We conclude the section by drawing insight from the numerical simulation on practical communication scenarios.
The network in Fig. \ref{fig:simsetup_a_b} well models a practical, LTE-style cellular networks in which RX~2 is a receiver located at the coverage edge among relay nodes while RX~1 and RX~3  are users served in a good network coverage area.
Since RX~2 is at the edge of two cells, a reasonable scenarios would be to assume $a$ small.
Also, since RX~1 and RX~3 are within the cell boundary, we expect $b$ to be small.

End users operating at the edge of coverage area of two or more nodes are a common and problematic scenario in cellular networks, especially in LTE-A networks.
While in GSM networks this issue was addressed by frequency reuse planning, in the predecessor of LTE-A, UMTS networks, such cases were dealt by so-called soft handovers, where the cell edge users were served by more than one cell simultaneously and selection combining was compensating for poor coverage and interference, decreasing the power consumption at each cell as well as the interference toward the users with good network coverage \cite{schinnenburg2003realization}.
Due to architectural changes and faster scheduling demands, which limits the cooperation among the base stations, soft handover is no more an available feature in LTE \cite{racz2007handover}, leaving the issue of users on the coverage borders an open problem.
The research needed to address this problem is ongoing for future LTE releases and several possible solutions were described in this work.
%
Solutions like fractional frequency reuse \cite{stolyar2008self, lei2007novel} are being considered but are not efficient as they limit the frequency spectrum allocation and,consequently, user throughputs (although making the frequency allocation more flexible).

%
%
%

Cooperative strategies, such as CoMP, represent the most promising way of optimizing the performance of coverage edge users.
Cooperation among serving nodes is becoming feasible for the next generation of wireless networks.
The new decentralized network architecture of LTE poses difficulties in cooperation taking place  among multiple base stations, due to delay issues.
On the other hand, the cooperation among the nodes served by a single base station is a more feasible and promising solution \cite{sawahashi2010coordinated}.

Considering these trends, the scenario under our consideration is indeed very relevant for practical applications.
%

From the numerical simulation of the CGRAS without rate-splitting in Sec. \ref{sec:No rate-splitting Scenario}, we conclude that cooperation among RNs  toward the cell edge user RX~2 is optimal.
%
%
In fact, regardless of the symmetric rate, in the region for small $a$ and small $b$ the optimal schemes are 14, 15 and 18 , all of which consider cooperation over $W_2$.

The drawback of cooperation is the tradeoff between relay link and access link power consumption and the resource allocation.
%
In particular, spectrum  availability at the relay link may become a bottleneck in practical systems, assuming the base station also serves some end users directly.
However, the common target in network planning is to minimize the coverage edge areas and for the coverage overlap among two and three cells not to exceed 30\% and 10\% respectively of the total covered area, making the cooperation for edge users less resource consuming from the relay link point of view.

%
Moreover, the simulation results show it is optimal set the codeword of the edge user as a bottom codeword for superposition,
which implies that it is optimal for the users in a better coverage of both relay nodes to decode the message transmitted to the edge user.

Similar conclusions can be drawn from the simulation with rate-splitting: key areas where cooperation is optimal are the areas with lower parameter $a$ values
in which  again the codeword for RX~2 acts as a bottom codeword for superposition and is therefore decoded by the other receivers.

\section{Conclusion}
\label{sec:Conclusion}
The relationship between cooperation and energy efficiency in relay-assisted downlink cellular system is studied through an information theoretical approach.
We consider in particular the case in which a base station communicates to the three receivers through the aid of two relays.
We allow for the base station to transmit the message of one receiver to multiple relay nodes which makes it possible for the relay nodes to partially coordinate their transmission.
%
This scenario idealizes LTE-style cellular networks in which relay nodes are used to improve the energy efficiency of the network.
We present upper and lower bounds to the overall energy consumption in this network which provide important insight on energy efficient transmission schemes for larger networks.
Our results clearly show that relay cooperation is necessary in attaining a higher power efficiency along with high overall network throughput.
In particular cooperation is necessary when transmitting toward users on the cell edge and which have low gain toward multiple relays.
We also present a new numerical simulation tool which generates a large number of achievable rate regions based on superposition coding, rate splitting and interference decoding.
With this tool we can consider a large number of transmission strategies and choose the most power efficient.
The power optimization of each achievable rate region can be performed in with low complexity and thus this approach can be easily extended to larger networks.

\section*{Acknowledgments}
The authors would like to thank Prof. Gerhard Kramer for the stimulating conversations and useful comments.

\bibliographystyle{IEEEtran}
\bibliography{literature,steBib1}

\end{document}